 \documentclass[final,5p,times,twocolumn,authoryear]{elsarticle}

\usepackage{amssymb}
\usepackage{lipsum}
\usepackage{url}
\usepackage{amsmath}
\usepackage{threeparttable} 
\usepackage{xcolor}
\usepackage{hyperref}
\usepackage[utf8]{inputenc}

\journal{High Energy Astrophysics}

\begin{document}

\begin{frontmatter}



\title{The origin of very-high-energy gamma-rays from GRB 221009A: implications for reverse shock proton synchrotron emission}


\author[first]{B. Theodore Zhang}
\ead{bing.zhang@yukawa.kyoto-u.ac.jp}
\author[second,third,fourth,first]{Kohta Murase}
\ead{murase@psu.edu}
\author[first]{Kunihito Ioka}
\author[fifth,sixth]{Bing Zhang}

\affiliation[first]{organization={Center for Gravitational Physics and Quantum Information, Yukawa Institute for Theoretical Physics, Kyoto University},
            addressline={}, 
            city={Kyoto},
            postcode={Kyoto 606-8502}, 
            state={},
            country={Japan}}

\affiliation[second]{organization={Department of Physics, The Pennsylvania State University},
            addressline={}, 
            city={University Park},
            postcode={PA 16802}, 
            state={},
            country={USA}
            }

            \affiliation[third]{organization={Department of Astronomy \& Astrophysics, The Pennsylvania State University},
            addressline={}, 
            city={University Park},
            postcode={PA 16802}, 
            state={},
            country={USA}}
            
            \affiliation[fourth]{organization={Center for Multimessenger Astrophysics, Institute for Gravitation and the Cosmos, The Pennsylvania State University},
            addressline={}, 
            city={University Park},
            postcode={PA 16802}, 
            state={},
            country={USA}}
            
            \affiliation[fifth]{organization={Nevada Center for Astrophysics, University of Nevada},
            addressline={}, 
            city={Las Vegas},
            postcode={NV 89154}, 
            state={},
            country={USA}}
            
                        \affiliation[sixth]{organization={Department of Physics and Astronomy, University of Nevada Las Vegas},
            addressline={}, 
            city={Las Vegas},
            postcode={NV 89154}, 
            state={},
            country={USA}}

\begin{abstract}
Recently, GRB 221009A, known as the brightest of all time (BOAT) GRB, has been observed across an astounding range of $\sim 18$ orders of magnitude in energy, spanning from radio to very-high-energy (VHE) bands.
Notably, the Large High Altitude Air Shower Observatory (LHAASO) recorded over $60000$ photons with energies exceeding $0.2\rm~TeV$, including the first-ever detection of photons above $10\rm~TeV$.
However, explaining the observed energy flux evolution in the VHE band alongside late-time multi-wavelength data poses a significant challenge.
Our approach involves a two-component structured jet model, consisting of a narrow core dominated by magnetic energy and a wide jet component dominated by matter. We show that the combination of the forward shock electron synchrotron self-Compton emission from the two-component structured jet and reverse shock proton synchrotron emission from the wide jet could account for both the energy flux and spectral evolution in the VHE band, and the early TeV lightcurve may be influenced by prompt photons which could explain the initial steep rising phase. We notice the arrival time of the highest energy photon $\sim$ 13 TeV
detected by LHAASO, especially a minor flare and spectral hardening occurring about $\sim$ 500 - 800 seconds after the trigger, is consistent with the emergence of the reverse shock proton synchrotron emission.
These findings imply that the GRB reverse shock may serve as a potential accelerator of ultra-high-energy cosmic rays, a hypothesis that could be tested through future multimessenger observations.
\end{abstract}




\end{frontmatter}




\section{Introduction}
\label{introduction}
The detection of very-high-energy (VHE) gamma rays (with energies ranging from 0.1 TeV to 100 TeV) originating from gamma ray bursts (GRBs) marks a significant advancement in our ability to explore the physics of GRBs and their associated radiative processes~\citep[See][for a review]{Meszaros:2006rc, Kumar:2014upa, zhangPhysicsGammaRayBursts2018}.
The initial detections in the VHE band included GRB 190114C observed by the Major Atmospheric Gamma Imaging Cherenkov (MAGIC) telescopes~\citep{MAGIC:2019lau, MAGIC:2019irs} and GRB 180720B observed by the High Energy Stereoscopic System (H.E.S.S.)~\citep{Abdalla:2019dlr}. 
Subsequently, VHE gamma rays from GRB 190829A~\citep{HESS:2021dbz} and GRB 201216C~\citep{2020GCN.29075....1B} were also reported. 
Furthermore, there are two sub-threshold events, namely GRB 160821B~\citep{MAGIC:2020ikk} and GRB 201015A~\citep{2020GCN.28659....1B}.
The four significantly detected VHE GRBs are long GRBs~\citep{Noda:2022hbo}, { and all} these detections align with the hypothesis that VHE gamma rays originate during the afterglow phase when the relativistic outflow enters the self-similar deceleration phase~\citep{Zhang:2019aia, Miceli:2022efx}.
However, it is worth noting that both MAGIC and H.E.S.S. operate as Imaging Atmospheric Cherenkov Telescopes (IACTs), characterized by limited fields of view and the necessity for longer slew times to perform follow-up observations.
Consequently, the detection of VHE gamma rays from the early afterglow, especially during the prompt phase, presents a formidable challenge.

On October 9, 2022, the Fermi
gamma ray Burst Monitor (Fermi-GBM) detected GRB 221009A at $T_0$ = 13:16:59.99 UT~\citep{Lesage:2023vvj}.
Because of { the highest fluence during the prompt phase}, GRB 221009A has been referred to as the brightest of all time (BOAT)~\citep{Burns:2023oxn}. 
Furthermore, this GRB was also captured by the Large High Altitude Air Shower Observatory (LHAASO), an extensive air shower detector designed for gamma ray observations~\citep{Ma:2022aau}.
A detailed analysis of LHAASO's Water Cherenkov Detector Array (WCDA) data unveiled more than $64000$ photons with energies ranging from 0.2 TeV to 7 TeV that can be attributed to GRB 221009A within the first $T - T_0\sim 3000$ seconds following the Fermi-GBM trigger at $T_0$~\citep{LHAASO:2023kyg}. 
Moreover, the LHAASO-KM2A detector also registered dozens of photons with energies exceeding $\sim 7$ TeV, with the highest recorded photon energy surpassing 10 TeV~\citep{LHAASO:2023lkv}.
Notably, { the observed flux ratio between TeV and MeV emission during the main burst phase is $\lesssim 2\times 10^{-5}$, possibly due to internal $\gamma \gamma$ absorption~\citep{LHAASO:2023kyg}.}
For the first time, we observed the rising phase of energy flux in the VHE band, which reached its peak flux at $T_0 + 244\rm~s$ and then gradually declined over time.
This smooth temporal profile of the TeV light curve aligns with its origin in the afterglow phase~\citep{LHAASO:2023kyg}.

Multiple studies lend support to the idea that VHE gamma rays originate during the afterglow phase. 
In this scenario, the standard synchrotron self-Compton (SSC) process plays a pivotal role in generating the observed VHE gamma rays~\citep{Meszaros:1994sd, Dermer:1999eh, Sari:2000zp, Zhang:2001az}.
In the SSC model, the same group of electrons responsible for synchrotron emission can upscatter these photons to higher energies by a factor of approximately $\gamma_e^2$, where $\gamma_e$ represents the electron Lorentz factor.
{ This SSC framework is consistent with the emissions observed in many VHE GRBs~\citep[e.g.,][]{MAGIC:2019lau, Abdalla:2019dlr, Wang:2019zbs, Asano:2020grw, Derishev:2021ivd, 2021ApJ...917...95Z, Huang:2022fnv}, although the simplistic one-zone model appears challenging for some VHE GRBs, such as
GRB 190829A~\citep[e.g.,][]{HESS:2021dbz, Salafia:2021njb, Sato:2022wec, Ren:2022icq, Khangulyan:2023akp, Sato:2022kup}}.
Nevertheless, it remains unclear whether the SSC process is the sole mechanism responsible for producing VHE gamma rays in GRBs. 
An alternative process that warrants consideration is the external inverse-Compton (EIC) mechanism, particularly in scenarios where the prompt emission coincides with the early afterglow phase~\citep{Wang:2006eq, Murase+10,  Toma:2009mw, murase_implications_2011, He:2011aa, Veres:2013dea, Murase:2017snw}.
Recent studies lend credence to the EIC scenario as a viable alternative for VHE gamma ray production, especially when considering the presence of a long-lasting central engine~\citep{Zhang:2020tem} and/or flares~\citep{Zhang:2020qbt}. 

VHE gamma ray production may also involve hadronic processes associated with the acceleration of cosmic ray protons and ions. 
In the realm of high-energy astrophysical sources featuring relativistic jets, extensive research has focused on proton synchrotron emission~\citep[e.g.,][]{Totani:1998zr, Zhang:2001az, Murase:2008mr, asano_prompt_2009} and photomeson production processes~\citep[e.g.,][]{Waxman:1999ai}.
However, the absence of high-energy neutrinos from GRBs imposes stringent constraints on the efficiency of the photomeson production process~\citep[e.g.,][]{Murase:2022vqf, Ai:2022kvd, Liu:2022mqe, Rudolph:2022dky,IceCube:2023rhf}.
Conversely, the radiative efficiency of the proton synchrotron process is considerably lower than that of electrons, necessitating the acceleration of protons to ultra-high-energies (UHE, exceeding $1 \rm~EeV$).
\cite{Isravel:2022glo} utilized synchrotron emission from UHE protons, accelerated within the external forward shock, to account for the observed VHE gamma rays in GRB 190114C, with the SSC component in a subordinate role. 
Similarly, \cite{Huang:2023dvb} examined UHE protons accelerated during the prompt emission phase and injected into the afterglow jet, demonstrating that proton synchrotron emission could dominate the early TeV afterglow of GRB190829A.
Inspired by the detection of $\gtrsim 10\rm~TeV$ photons from GRB 221009A, \cite{Zhang:2022lff} put forward the idea that reverse shock proton synchrotron emission could be a significant contributor to VHE gamma ray production in the case of GRB 221009A~\citep[See][for the external forward shock model.]{Isravel:2023thi}.

Studying the production of VHE gamma rays in detail is crucial, especially considering the remarkable TeV flux light curve observed by LHAASO and the wealth of multiwavelength data available. 
In their study~\cite{LHAASO:2023kyg}, they conducted comprehensive numerical modeling of the TeV light curve, focusing on the time window covered by LHAASO's observations.
They found that the early TeV emission can be adequately modeled by SSC emission from a narrow jet. On the other hand,  the late-time radio to GeV data has to be attributed to a second jet~\citep{Sato:2022kup, Zheng:2023crd, 2023arXiv231114180Z}.  
While the TeV light curve suggests a preference for a constant external medium density distribution\footnote{{Note the wind medium case can not be excluded~\citep{Khangulyan:2023srq}.}}~\citep{LHAASO:2023kyg, Sato:2022kup}, the late-time radio afterglow aligns with the propagation in a wind medium~\citep{Ren:2022icq, Gill:2023ijr}.
In our earlier work of ~\cite{Zhang:2022lff}, we focused on the moment when the reverse shock has fully traversed the ejecta, marking the point at which the reverse shock emission attains its peak luminosity.
Interestingly, our analysis revealed that proton synchrotron emission exhibits hard spectra, and the maximum energy of these emissions can surpass $10\rm~TeV$.
However, \cite{Zhang:2022lff} came out before the availability of the LHAASO data and without considering the realistic dynamical evolution of the jet, so it is reasonable to perform detailed calculations of the reverse shock proton synchrotron model in this work.

In this study, we conduct comprehensive numerical modeling of VHE gamma ray production and multiwavelength light curves using the standard reverse-forward external shock model with a structured jet. 
We investigate various radiative processes involving nonthermal electrons and protons from both the shocked ejecta and the shocked external medium.
Our paper is organized as follows: 
In Section~\ref{sec:model}, we provide detailed information about the numerical modeling of the reverse-forward shock emission within the structured jet.
In Section~\ref{sec:result}, we present our findings and compare them to the observed multi-wavelength light curves and energy spectra from GRB 221009A.
In Section~\ref{sec:discuss}, we discuss the implications of our results.
Finally, we summarize our study in Section~\ref{sec:summary}.

In this study, we employ centimetre-gram-second system units and express quantities as $Q=Q_x 10^{x}$.
We denote photon energy in the observer frame as $E$ and in the comoving frame as $\varepsilon$.

\section{Physical model}\label{sec:model}
\subsection{Structured jet}
Structured jets in the context of GRBs have been extensively studied in previous research~\citep{1998ApJ...499..301M, Rossi:2001pk, Zhang:2001qt, 2003ApJ...591.1075K, 2004ApJ...601L.119Z}.
The detection of the gravitational wave event GW 170817~\citep{LIGOScientific:2017vwq} and the associated short GRB 170817A~\citep{LIGOScientific:2017ync} provides a unique opportunity to explore the characteristics of relativistic jets.
The gradually rising radio to X-ray afterglow observations support the off-axis structured jet model~\citep{PhysRevLett.120.241103, 2018ApJ...856L..18M, 2018MNRAS.478L..18T, 2018PTEP.2018d3E02I}.
Structured jet formation is a natural outcome when the jet interacts with the stellar envelope~\citep{Gottlieb:2020mmk, Gottlieb:2020raq}.
Interestingly, the fitting of the TeV light curve suggests a preference for emission from a narrow jet, which may correspond to the core of a structured jet~\citep{LHAASO:2023kyg}.
Furthermore, the observed X-ray decay rate is shallower than the predicted post-jet-break index for a standard top-hat jet, suggesting that relativistic jets likely exhibit angular structure in their energy profiles~\citep{OConnor:2023ieu, Gill:2023ijr}.

\begin{figure}
    \centering
    \includegraphics[width=0.5\textwidth]{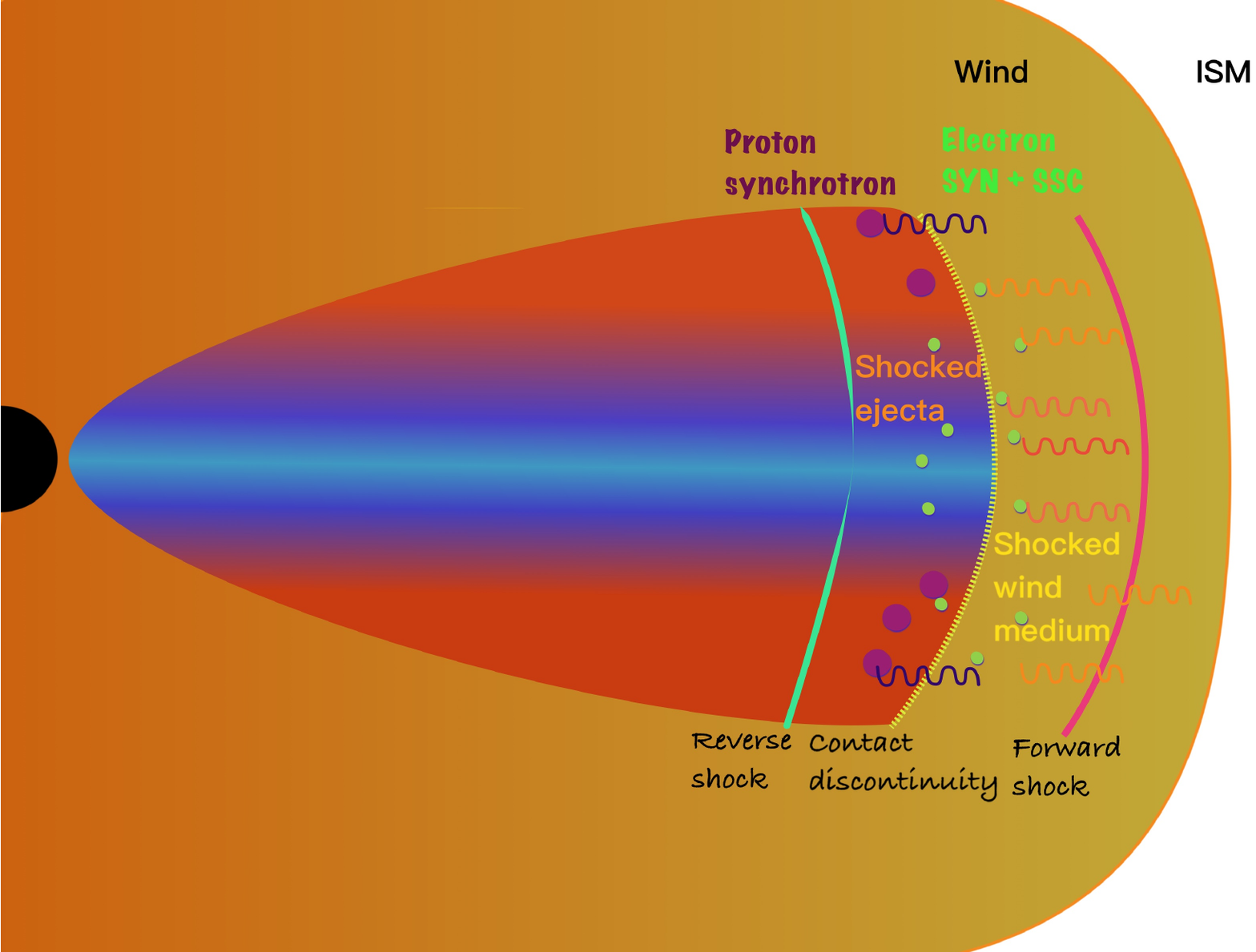}
    \caption{A schematic depiction of the structured jet, featuring a magnetic-energy-dominated narrow core (illustrated with blue color) and a matter-dominated wide jet component (illustrated with red color).}
    \label{fig:jet}
\end{figure}

In this study, we present an illustrative representation of a structured jet, which consists of a magnetic-dominated narrow core and a matter-dominated wide jet component, as depicted in Fig.~\ref{fig:jet}. The arguments in support of such a two-component jet with different jet compositions have been presented in detail in \cite{2023arXiv231114180Z}.
The magnetic-dominated narrow core is represented as a top-hat jet with a uniform energy and velocity distribution. 
Conversely, the matter-dominated wide jet component is modeled as a power-law structured jet.

We define $\mathcal{E}(\theta)$ as the angle-dependent isotropic-equivalent energy of a given structured jet.
For the power-law structured jet, we have~\citep[e.g.][]{Takahashi:2022xxa}
\begin{equation}
    \mathcal{E}(\theta) = \mathcal{E}_k \left[1 + \left(\frac{\theta}{\theta_c}\right)^2\right]^{-a/2} \ (\theta < \theta_j),
\end{equation}
where $\mathcal{E}_{k}$ represents the isotropic-equivalent kinetic energy, $\theta_c$ is the narrow cone angle, $a$ is the power-law index, and $\theta_j$ is the jet half-opening angle.
The distribution of the Lorentz factor is described as
\begin{equation}
    \Gamma(\theta) = \Gamma_0 \left[1 + \left(\frac{\theta}{\theta_c}\right)^2\right]^{-a/2} \ (\theta < \theta_j),
\end{equation}
where $\Gamma_0$ is the { initial bulk Lorentz factor of the ejecta}.
In the case of a uniform top-hat jet, $a = 0$.

\subsection{Ambient medium}
Late-time modeling of the radio afterglow from GRB 221009A suggests that the external medium follows a wind profile, which decreases outwards as $n_{\rm ex} (R) \propto R^{-2}$~\citep[e.g.,][]{Ren:2022icq, Gill:2022erf}.
However, the { slow rise in VHE band energy flux, characterized by a slope of approximately 1.8, suggests a preference for either a constant ambient density distribution near the progenitor or energy injection in the wind medium}~\citep{LHAASO:2023kyg}.
The transition from a constant ambient density to a wind medium leads to a change in the slope of the multiwavelength light curve.
It's important to note that the early afterglow light curve may be shaped by $\gamma \gamma$ attenuation from external prompt photons~\citep[e.g.,][]{Zhang:2022lff, Khangulyan:2023srq}.

\begin{figure}
    \centering
    \includegraphics[width=0.5\textwidth]{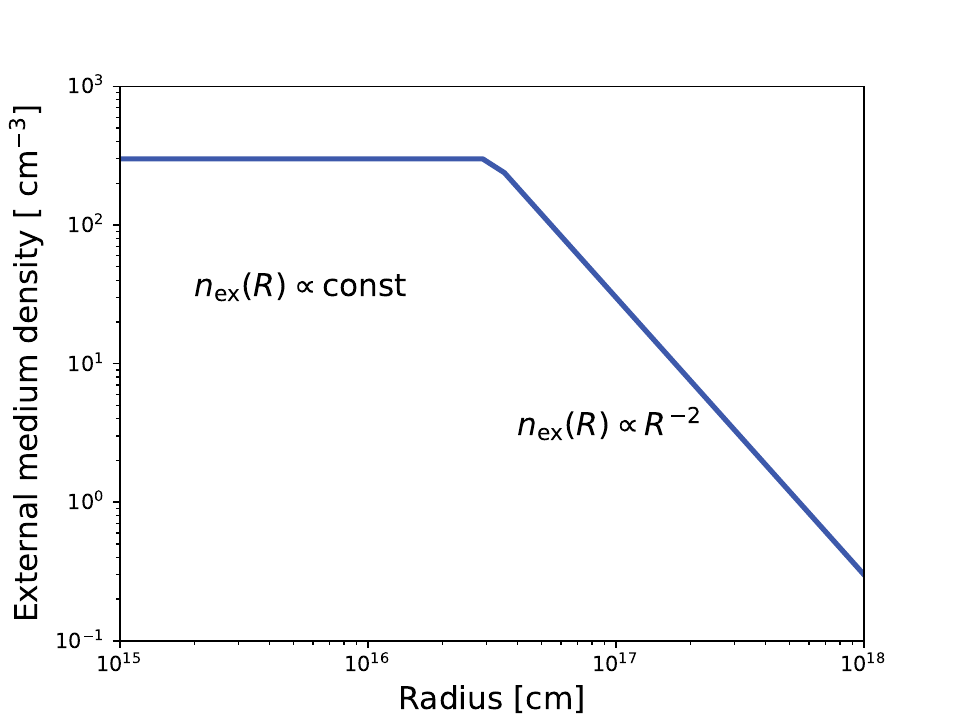}
    \caption{The number density distribution of the external medium as a function of radius from the central engine in the laboratory frame. Here, we adopt $n_{\rm ex, 0} = { 300}\rm~cm^{-3}$ and $A_* = 1$.}
    \label{fig:external-density}
\end{figure}

In our study, we consider a scenario where the external medium is predominately composed of a stellar wind. 
However, in the inner region near the progenitor star, the density profile of the external medium could be influenced by pre-explosion burst activities, resulting in a constant density profile. 
We model the external medium using the following formula,
\begin{equation}
n_{\rm ex} (R) = {\rm min} [n_{\rm ex, 0}, A R^{-k}],
\end{equation}
where $A = 3 \times 10^{35} A_* {\rm cm}^{-1}$ { and $k=2$ for a stellar wind medium. The parameter $A_*$ } represents the ratio of mass-loss rate to wind speed, normalized to $10^{-5} M_\odot \ {\rm yr}^{-1} / 10^3 {\rm~km \ s^{-1}}$.
The number density distribution of the external medium is illustrated in Fig.~\ref{fig:external-density}.

\subsection{Particle acceleration and radiative process}
As the relativistic ejecta interacts with the external medium, two shocks form: the forward shock, advancing into the ambient external medium, and the reverse shock, moving backward to decelerate the outflow.
In this context, we denote the downstream internal energy density (pressure) of the shocked wind medium as $e_2$ ($p_2$) and the downstream internal energy density (pressure) of the shocked ejecta as $e_3$ ($p_3$), as depicted in Fig.~\ref{fig:jet}.
Given pressure equilibrium across the contact discontinuity, described by $p_2 = p_3$, we find the relation 
\begin{equation}
    (\hat{\gamma}_2 - 1) e_2 = (\hat{\gamma}_3 - 1) e_3,
\end{equation}
where $\hat{\gamma}_2$ and $\hat{\gamma}_3$ represent the adiabatic indices in shocked ejecta and shocked external medium, respectively.
The adiabatic indices can be expressed as
\begin{equation}
    \hat{\gamma} \approx \frac{4\bar{\gamma} + 1}{3\bar{\gamma}},
\end{equation}
where $\bar{\gamma}$ represents the average Lorentz factor of particles.
For non-relativistic case, $\hat{\gamma} = 5/3$, whereas for relativistic case, $\hat{\gamma} = 4/3$~\citep{zhangPhysicsGammaRayBursts2018}.

The amplification of magnetic field strength in the downstream region of the reverse-forward shock system remains unclear~\citep[e.g.,][]{Medvedev:1999tu}. 
We parameterize a fraction $\epsilon_B$ ($\epsilon_B^r$) of the downstream internal energy density of the shocked external matter (ejecta) being converted into magnetic field energy density, $u_B$ ($u_B^r$). Note we use the superscript $r$ to represent the parameters in the shocked ejecta.
Both electrons and protons (or ions) can be accelerated as they traverse the shock under the diffusive shock acceleration mechanism (DSA)~\citep[e.g.,][]{Sironi:2015oza}. 
These accelerated particles exhibit a power-law distribution, with their maximum energy determined by the balance between the acceleration timescale and various energy loss timescales~\citep[e.g.,][]{Asano:2020grw}.
The electron and proton spectral index are denoted as $s_e^r$ and $s_p^r$, respectively.
However, the energy content of these nonthermal particles remains unknown. 
We assume a fraction $\epsilon_e$($\epsilon_e^r$) of the downstream internal energy density goes into nonthermal electrons $u_e$ ($u_e^r$), and another fraction $\epsilon_p$($\epsilon_p^r$) of the downstream internal energy density is allocated to nonthermal protons $u_p$($u_p^r$). In addition, we assume a number fraction $f_e$($f_e^r$) of electrons can be accelerated.
These microphysical parameters can potentially be constrained by fitting multi-wavelength afterglow light curves.

Studies have shown that accelerating protons to ultra-high energies (UHE) is a slow process for ultra-relativistic external forward shocks due to the relatively weak magnetic field in the surrounding medium~\citep[e.g.,][]{Gallant:1998uq, Murase:2008mr, Sironi:2015oza}. 
Conversely, the reverse shock has been proposed as an alternative acceleration site for accelerating UHECRs~\citep[e.g.,][]{Waxman:1999ai, Murase:2007yt, Murase:2008mr, Zhang:2017moz}.
Note we only consider reverse shock emission from the matter-dominated wide jet, because the reverse shock is suppressed in the magnetic-dominated narrow core~\cite{Zhang:2004ie}.
Unlike the forward shock, the reverse shock could be either nonrelativistic or only mildly relativistic as shown in Eq.~\ref{eq:gamma34} ~\citep{sari_hydrodynamic_1995, sari_hydrodynamics_1997}.
Building upon the concept proposed in Ref.~\cite{Zhang:2022lff}, we explore the acceleration of UHE protons by the reverse shock and investigate proton synchrotron emission while considering the dynamical evolution of the reverse shock.
Ref.~\cite{Zhang:2022lff} also considered reverse shock EIC radiation. This component is relevant for explaining the 0.1-10 GeV gamma rays measured by \textit{Fermi}-LAT, which is not the focus of this work. However, one should keep in mind that proton synchrotron emission studied in this work works self-consistently with the EIC model presented in Ref.~\cite{Zhang:2022lff}.

\subsection{Numerical methods}
The calculations in this work are based on the Astrophysical Multimessenger Emission Simulator (AMES) code that includes a module to numerically model multi-wavelength emission from GRB afterglows\footnote{\url{https://github.com/pegasuskmurase/AMES-GRBAfterglow}}. 
We expand the GRB-afterglow module to treat emissions from not only an external forward shock but also a reverse shock.   
For a comprehensive understanding of the dynamic evolution of the reverse-forward shock system and the associated radiative processes, see Appendix~\ref{eq:appendix_numerical}. Below, we describe the general calculation processes.

The observed flux at a given observation time $T$ could be determined by integrating over the equal-arrival-time-surface (EATS)~\citep[e.g.,][]{zhangPhysicsGammaRayBursts2018}.
In our calculations, we adopt the thin-shell limit~\citep{vanEerten:2010zh, Ryan:2019fhz, Takahashi:2019otc, Takahashi:2022xxa}.
The observed flux can be estimated as follows~\citep[e.g.][]{Takahashi:2022xxa}
\begin{align}\label{eq:flux-EATS}
    F_E (T) &= \frac{{ 1 + z}}{d_L^2} \int_0^{\theta_j} d\theta {\rm sin}\theta \int_0^{2\pi} d\phi \frac{R^2 |\mu - \beta_{\rm sh}|}{1-\mu \beta_{\rm sh}} \nonumber \\ &\times \frac{j_{E_z}}{\alpha_{E_z}} (1 - e^{-\tau_{E_z}})|_{\hat{t} = T_z + \mu R / c}.
\end{align}
Here, { $F_E = dF/dE$ is the specific flux, $T$ represents the observation time, $z$ is the redshift, $T_z = T / (1 +z)$, $d_L$ stands for the luminosity distance of the source, $\theta$ denotes the polar angle measured from the jet axis, $\phi$ represents the azimuth angle, $j_{E_z}$ refers to the emissivity measured in the laboratory frame, $\alpha_{E_z}$ to the absorption factor measured in the laboratory frame, $\tau_{E_z}$ to the corresponding optical depth measured in the laboratory frame, and $\hat{t}$ represents the laboratory time.} 
The term $\mu$ is given by
\begin{equation}
\mu = {\rm sin}\theta {\rm sin} \theta_v {\rm cos} \phi + {\rm cos} \theta {\rm cos} \theta_v,
\end{equation}
which determines the cosine of the angle formed by the radial vector in the ($\theta,\phi$) direction and the line-of-sight direction on the x-z plane.
Here, $\theta_v$ corresponds to the viewing angle measured from the jet axis. 
The shock front velocity divided by the speed of light is denoted as $\beta_{\rm sh} = \sqrt{1 - \Gamma_{\rm sh}^{-2}}$, where $\Gamma_{\rm sh} = \sqrt{2} \Gamma$. Here, $\Gamma$ represents the Lorentz factor of the blastwave fluid immediately behind the shock front.
In our calculations, we set $T = 0$ as the arrival time when a photon is emitted at the origin at the laboratory time $\hat{t} = 0$.
The width of the shocked shell in the laboratory frame is given by
\begin{equation}\label{eq:width}
    w = \frac{R}{4 (3 - k) \Gamma^2}.
\end{equation}
The emissivity measured in the laboratory frame is
\begin{equation}
    j_{E_z} = \delta^2 j_{\varepsilon} = \frac{j_{\varepsilon}}{\Gamma^2 (1 - \beta \mu)^2},
\end{equation}
where $j_{\varepsilon}$ is the comoving frame emissivity per unit volume per second per solid angle, $\delta$ is the Doppler factor, and 
\begin{equation}
E_z = E (1+z) = \delta \varepsilon = \frac{\varepsilon}{\Gamma (1 - \beta \mu)},    
\end{equation}
is the photon energy in the laboratory frame.
The absorption coefficient in the laboratory frame is given by
\begin{equation}
    \alpha_{E_z} = \Gamma (1 - \beta \mu) \alpha_{\varepsilon},
\end{equation}
where $\alpha_{\varepsilon}$ is measured absorption coefficient in the comoving frame.
The corresponding optical depth can be derived via integration over the line of sight,
\begin{equation}
    \tau_{E_z} = \int \alpha_{E_z} ds \approx \alpha_{E_z} \Delta s.
\end{equation}
where $\Delta s$ represents a finite segment along a ray  traveling the shocked region
where emitted photons reach the observer at time $T$.
The value of $\Delta s$ can be expressed as 
\begin{equation}
    \Delta s = \frac{R}{4 (3 - k) \Gamma^2 |\mu - \beta_{\rm sh}|},
\end{equation}
as shown in~\cite{Takahashi:2022xxa}.
In the optically thin limit, Eq.~\ref{eq:flux-EATS} can be simplified as 
\begin{align}\label{eq:flux-EATS-thin}
    F_E (T) &= \frac{1+z}{d_L^2} \int_0^{\theta_j} d\theta {\rm sin}\theta \int_0^{2\pi} d\phi R^2 \Delta R j_{E_z}|_{t = T_z + \mu R / c},
\end{align}
where
\begin{equation}
    \Delta R = \frac{R}{4 (3 - k) \Gamma^2 (1 - \mu \beta_{\rm sh})}
\end{equation}
represents the radial integration path that contributes to the observed flux at a given observation time $T$.

In this work, we primarily focus on two key attenuation processes: synchrotron self-absorption (SSA), crucial for understanding low-frequency radio emission, and $\gamma \gamma$ pair production, which affects the escape of VHE gamma rays.
In the latter situation, the primary targets for interaction are the nearby low-energy synchrotron photons emitted by nonthermal electrons. 
The absorption coefficient is estimated as
\begin{equation}
    \alpha_{\varepsilon} \approx \frac{1}{c t_{\gamma \gamma}},
\end{equation}
where $t_{\gamma \gamma}$ is the comoving frame interaction timescale. 
{ Note that we only consider $\gamma \gamma$ attenuation by the target photons produced in the same shell. It is possible that interactions with target photons from other shells could also contribute to attenuation. However, the density of the target photons from these other shells would be diluted due to expansion after crossing the emitting shell.}
During our numerical calculation process, we initially overlook the $\gamma \gamma$ attenuation caused by prompt photons. 
However, we later account for this effect in a subsequent post-processing step, which we detail in the next section.
Note in our calculations, we perform calculations in the momentum space, where the \textit{Deep Newtonian} limit is avoided~\citep{Huang:2003nw, Sironi:2013tva, Wei:2023rmr, Ryan:2023pzk}.

\section{VHE gamma rays from GRB afterglow}
\label{sec:result}

\subsection{Early afterglow in the VHE band}

\begin{figure*}
    \centering
    \includegraphics[width=1.\textwidth]{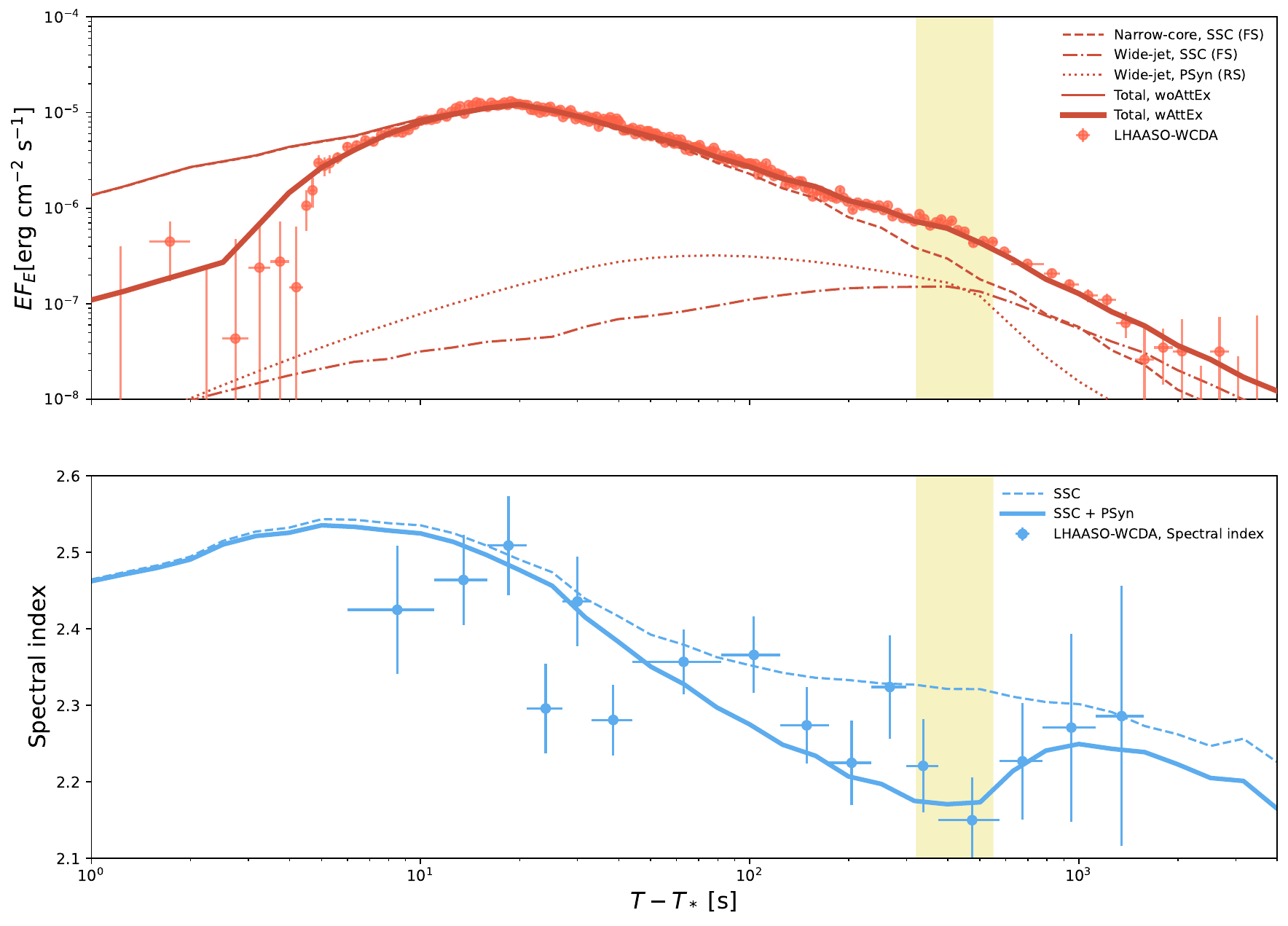}
    \caption{Predicted energy flux (upper panel) and spectral index (lower panel) evolution in the VHE band within the two-component structured jet model. 
    The yellow band indicates the region where there is a small flare.
    }
    \label{fig:lightcurve-VHE}
\end{figure*}

Our result is shown in Fig.~\ref{fig:lightcurve-VHE}, calculated using the updated GRB module of the \textsc{AMES} code described above.
In the upper panel of Fig.~\ref{fig:lightcurve-VHE}, we compare numerical results with energy flux measurements by LHAASO-WCDA integrated over the 0.3 TeV to 5 TeV energy range~\citep{LHAASO:2023kyg}, where the corresponding parameters are listed in Table~\ref{tab:parameter}.
There is a clear agreement between the predicted light curve (solid red line) and the measured energy flux (red dots) for about $T - T_* \sim 3000$ seconds after the GBM trigger, where $T_* = 226\rm~s$ is the reference time at the onset of the main component { of the afterglow light curves} ~\citep{LHAASO:2023kyg}.

However, our model's light curve shows an inconsistency with the measured one at early times, $T - T_* \lesssim 10\rm~s$, see solid thin lines. 
To be specific, the LHAASO-WCDA light curve exhibits an initial rapid rise with a power-law index of approximately $14.9$, followed by a slower ascent with a power-law index of roughly $1.82$, leading up to its peak value~\citep{LHAASO:2023kyg}.
In contrast, our predicted early light curve is less steep than the observed value, see solid red line. 
One potential explanation for this discrepancy could be the rapid initial rise being influenced by $\gamma\gamma$ attenuation due to the intense prompt photons~\cite[e.g.,][]{Zhang:2022lff, Khangulyan:2023srq}. 
Note a different $T_*$ will give a different rising slope, which could change the above interpretation.

The comoving frame energy density of the prompt photons at the external shocked region can be estimated as 
\begin{equation}
   u_{\rm GRB\gamma} \approx L_{\rm GRB\gamma}^{\rm iso}/ 4\pi R^2 \Gamma^2 c, 
\end{equation}
where $L_{\rm GRB\gamma}^{\rm iso}$ represents the isotropic-equivalent gamma ray luminosity.
The optical depth to the two-photon annihilation can be approximated as follows~\citep{Murase:2022vqf}:
\begin{align}\label{eq:optical-depth}
\tau_{\gamma\gamma}^{\rm prompt} &\approx \sigma_{\gamma \gamma} \frac{R}{\Gamma} \frac{\Gamma u_{\rm GRB\gamma}}{E_{\rm pk}} \nonumber \\ &\simeq 220\frac{\eta_{\gamma\gamma,-1}L_{\rm GRB\gamma, 53.5}^{\rm iso}}{R_{16}\Gamma_{2}^2 E_{\rm pk, \rm MeV}}
\begin{cases} (E_\gamma / \tilde{E}_{\gamma, b})^{\beta - 1}, E_\gamma < \tilde{E}_{\gamma, b} \\ (E_\gamma / \tilde{E}_{\gamma, b})^{\alpha - 1}, E_\gamma > \tilde{E}_{\gamma, b}
\end{cases}.
\end{align}
Here, $\sigma_{\gamma \gamma} \sim \eta_{\gamma \gamma} \sigma_T$ is the $\gamma \gamma$ pair production cross section~\citep{1987MNRAS.227..403S}, $\sigma_T$ is the Thomson cross section, $\alpha$ and $\beta$ represent the spectral indices of the prompt emission, $\tilde{E}_{\gamma, b} \approx \Gamma^2 m_e^2 c^4 / E_{\rm pk} \simeq 2.6~\Gamma_{2}^2 E_{\rm pk, \rm MeV}^{-1}\rm~GeV$ denotes the typical energy of high-energy $\gamma$-rays that interact with target photons of peak energy $E_{\rm pk}$ in the observer frame~\citep{Zhang:2022lff}.

Please note that we are disregarding the anisotropic scattering effect of the beamed prompt photons.
In the comoving frame of the external shock region, the direction of VHE gamma rays is isotropically distributed. Despite the monodirectional distribution of prompt target photons, those moving at a large angle relative to the direction of target photons may undergo annihilation. For a distant observer, most of the radiation arrives from the forward direction with $\theta^\prime \lesssim \pi / 2$ in the comoving frame, corresponding to $\theta \lesssim 1 / \Gamma$ in the observer frame. 
The anisotropic effect on the attenuation of gamma rays can be estimated 
via integration over $\mu^\prime$ in the comoving frame: 
\begin{align}
f_{\rm att} &= \int_0^{1} d\mu^\prime {\rm exp}\left[-\int_{\frac{2 m_e^2 c^4}{\varepsilon^\prime (1-\mu^\prime)}} d\varepsilon^\prime_t \frac{dn}{d\varepsilon^\prime_t}  (1-\mu^\prime) \sigma_{\gamma\gamma}(\varepsilon^\prime, \varepsilon^\prime_t, \mu^\prime) \frac{R}{\Gamma}\right] \nonumber \\ 
&\approx
\int_0^{\mu^\prime_{\rm th}} d\mu^\prime {\rm exp}[-(1-\mu^\prime)\tau_{\gamma\gamma}^{\rm prompt}] + \int_{\mu^\prime_{\rm th}}^{1} d\mu^\prime \nonumber \\ &\approx \frac{{\rm exp}[-\tau_{\gamma\gamma}^{\rm prompt}(1-\mu^\prime_{\rm th})] - {\rm exp}[-\tau_{\gamma\gamma}^{\rm prompt}]}{\tau_{\gamma\gamma}^{\rm prompt}} + (1 - \mu^\prime_{\rm th}), 
\end{align}
where $\tau_{\gamma\gamma}^{\rm prompt}$ is the optical depth calculated in Eq.~\ref{eq:optical-depth}, $\varepsilon^\prime$ represents the energy of the gamma rays, $\varepsilon^\prime_t$ denotes the target photon energy, $dn/d\varepsilon^\prime_t$ is the differential target photon number density, $\mu^\prime$ is the cosine of the angle between the direction of emitted photons and the direction of prompt target photons in the shock comoving frame and $R/\Gamma$ represents the comoving frame shock width. Here, $\mu^\prime_{\rm th}$ is the cosine of a threshold angle when $E_{\rm pk}/\Gamma \sim \frac{2 m_e^2 c^4}{\varepsilon^\prime (1-\mu^\prime_{\rm th})}$, which depends on the energy of gamma rays and target photon energy.
For smaller angles with $\mu^\prime > \mu^\prime_{\rm th}$, we simply assume that gamma rays can escape freely, considering the threshold energy for two-photon pair annihilation is larger than the peak energy of the prompt emission.
For TeV gamma rays with comoving frame energy $\varepsilon^\prime \sim 10\rm~GeV$ and $E_{\rm pk}/\Gamma \sim 10\rm~keV$, we derive $1 - \mu^\prime_{\rm th} = 0.005 \lesssim 1 / \tau_{\gamma\gamma}^{\rm prompt}$. Note the value of $\mu^\prime_{\rm th}$ is sensitive to both gamma ray energy and Lorentz factor.
This demonstrates that attenuation by anisotropic prompt photons leads to a suppression factor proportional to the inverse of the optical depth when $1 / \tau_{\gamma\gamma}^{\rm prompt} \gtrsim 1-\mu^\prime_{\rm th}$ and $\tau_{\gamma\gamma}^{\rm prompt} > 1$, especially for TeV gamma rays. 
 
During the early phase, the jet is in the coasting phase with a Lorentz factor of approximately $\Gamma \approx \Gamma_0$, and the radius of the external shock is approximately $R \approx 2\Gamma_0^2 c (T - T_*)$.
For illustrative purposes, we've parameterized the time evolution of the optical depth as 
\begin{equation}
    \tau_{\gamma\gamma}^{\rm prompt} = \tau_{\gamma\gamma, 0} \left(\frac{T - T_*}{1\rm~s}\right)^{-s_\tau},
\end{equation}
where $\tau_{\gamma\gamma, 0} \simeq {\rm 180}$ and $s_\tau = 3$. 
The value of $\tau_{\gamma\gamma, 0}$ is estimated using the parameters $L_{\rm GRB\gamma} = 10^{54}\rm~erg~s^{-1}$, $R = 10^{15.4}\rm~cm$ and $\Gamma = 400$.
In this context, we assume that $L_{\rm GRB\gamma}^{\rm iso} \propto (T - T_*)^{-2}$ and that $R \propto (T - T_*)$ during the early afterglow phase.
The prompt emission flux in the 20 keV - 10 MeV range is $1.62\times 10^{-2}\rm~erg~cm^{-2}~s^{-1}$ during the time interval $T_0 + [225.024-233.216]\rm~s$ and $3.15\times 10^{-4}\rm~erg~cm^{-2}~s^{-1}$ during the time interval $T_0 + [241.408-249.600]\rm~s$~\citep{Frederiks:2023bxg}. 
The energy flux decreases by two orders of magnitude when the VHE gamma ray light curve reaches its peak.
The observed flux in the VHE band, after accounting for the $\gamma\gamma$ attenuation by the prompt photons, can be obtained as, 
\begin{equation}\label{eq:atten_prompt}
    E F_E = E F_E|_{\rm woAttEx} f_{\rm att},
\end{equation}
where $E F_E|_{\rm woAttEx}$ is the flux without considering the attenuation by external prompt photons.
The attenuation factor for gamma rays with observed energies exceeding 300 GeV is $f_{\rm att} \sim 0.08$, gradually increasing to unity at later times.
Therefore, we attribute the rapid increase in early TeV afterglow data to the effect of $\gamma\gamma$ attenuation by the prompt photons, see the solid thick line in Fig.~\ref{fig:lightcurve-VHE}, which is consistent with the observed data and aligns with the argument presented in~\citep{Khangulyan:2023srq, Shen:2023qxp}.
Note that the reference time $T_*$ is uncertain, and a larger $T_*$ results in a shallower slope in the early light curve.

The energy flux in the VHE band peaks around $T - T_* \sim 18\rm~s$, which is consistent with the jet's duration, approximately $\delta T_{\rm ej} \sim 10\rm~s$.
Following the peak, the energy flux decreases over time, displaying a power-law spectral index of about $1.1$, consistent with the findings reported by~\cite{LHAASO:2023kyg}. 
It was also noted in~\cite{LHAASO:2023kyg} that the sharp decline in flux around $T-T_* \sim 670~\rm~s$ aligns with the jet break, resulting in an energy flux decreases with a power-law index of approximately $2.2$.
In our study, we propose an alternative explanation. 
We suggest that the steep drop in the light curve is caused by the emission from the wide jet (depicted as a red dotted-dash curve), while the jet break in the emission from the narrow core occurred earlier.
Furthermore, there's a small flare around $T - T_* = [320, 550]\rm~s$ as reported in~\citep{LHAASO:2023kyg}.
In our model, the emergence of reverse shock proton synchrotron emission (illustrated as a red dotted curve) appears to coincide with this observed flare.

\begin{table}
\caption{Physical parameters adopted in this work.}
\begin{threeparttable}[b]
    \centering
    \begin{tabular}{lcc} 
    Parameter &  Narrow-core  & Wide-Jet \\
    \hline
      $\mathcal{E}_k [\rm erg]$ & $4\times 10^{55}$ &  ${ 3}\times 10^{55}$ \\
      $\Gamma_0$   & $430$ & $400$ \\
      $n_{\rm ex, 0} [\rm cm^{-3}]$ & ${ 300}$ & ${ 300}$\\
      $A_*$ & 1 & 1 \\
      $k$ & 2 & 2 \\
      $\delta T_{\rm ej} [\rm s]$ & $10$ & $400 (3000)$\tnote{ } \\
      $\theta_j$ [deg] & { 0.12} & 5.7 \\
      $\theta_c [\rm deg]$ & - & 0.0005 \\
      $a$ & -  & 2 \\
      $\theta_v [\rm deg]$ & 0 & 0 \\
      \hline
      $\epsilon_B$ & { $ 3\times 10^{-4}$} & $1.5\times 10^{-4}$ \\
      $\epsilon_e$ & ${0.008}$ & 0.0022 \\
      $f_e$ & ${0.25}$ & 0.008 \\
      $s_e$ & $2.3$ & $2.3$ \\
      \hline
      $\epsilon_B^r$ & - & 0.5\tnote{ } \\
      $\epsilon_e^r$ & - & $0.002$ \\
      $f_e^r$ & - & $0.008$ \\
      $s_e^r$ & - & $2.4$ \\
      \hline
      $\epsilon_p^r$ & - & $0.25$ \\
      $s_p^r$ & - & $2$ \\
    \hline
    \end{tabular}
    \label{tab:parameter}
     \begin{tablenotes}
       \item [a] $\delta T_{\rm ej} = 3000\rm~s$ when $\theta_j > 1.1$ deg.
       \item [b] $\epsilon_B^r = 0.1$ when $\theta_j > 1.1$ deg.
     \end{tablenotes}
\end{threeparttable}
\end{table}

\subsection{Spectral hardening and reverse shock proton synchrotron emission}

\begin{figure*}
    \centering
    \includegraphics[width=0.45\textwidth]{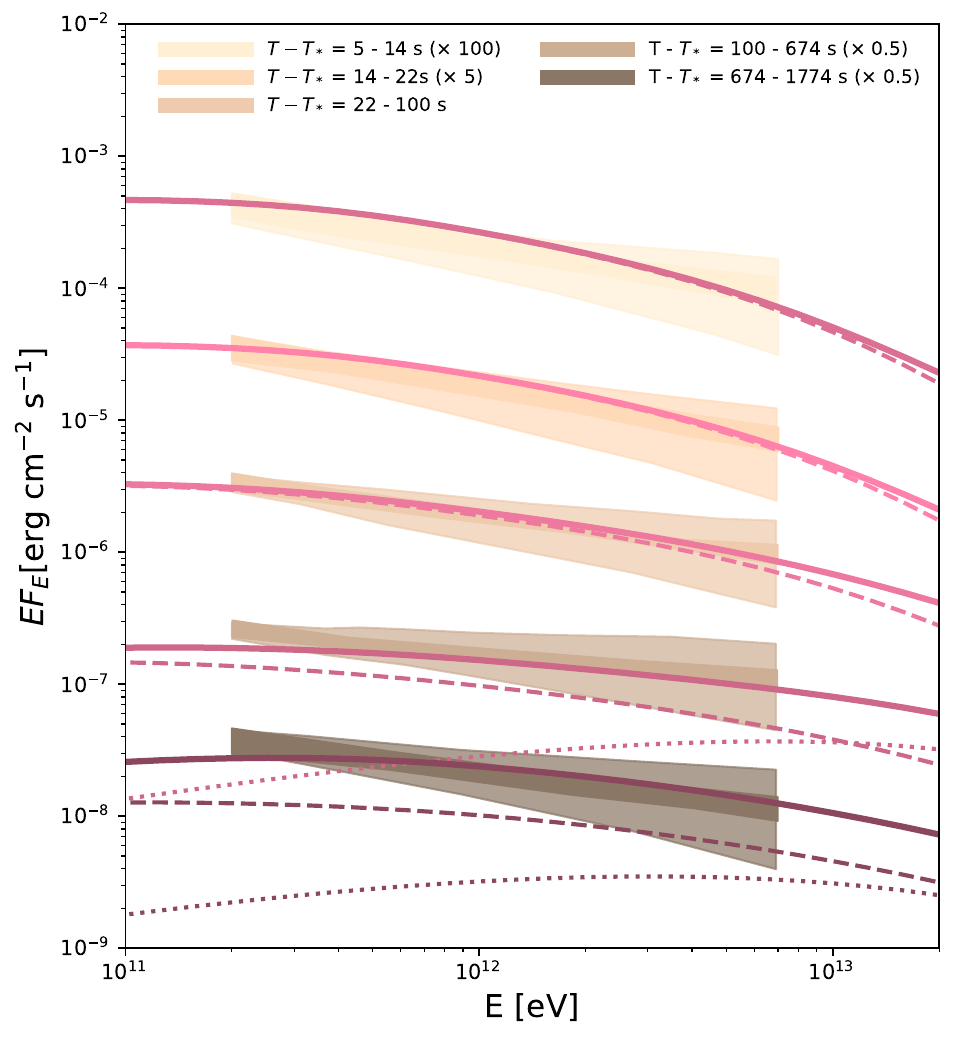}
    \includegraphics[width=0.45\textwidth]{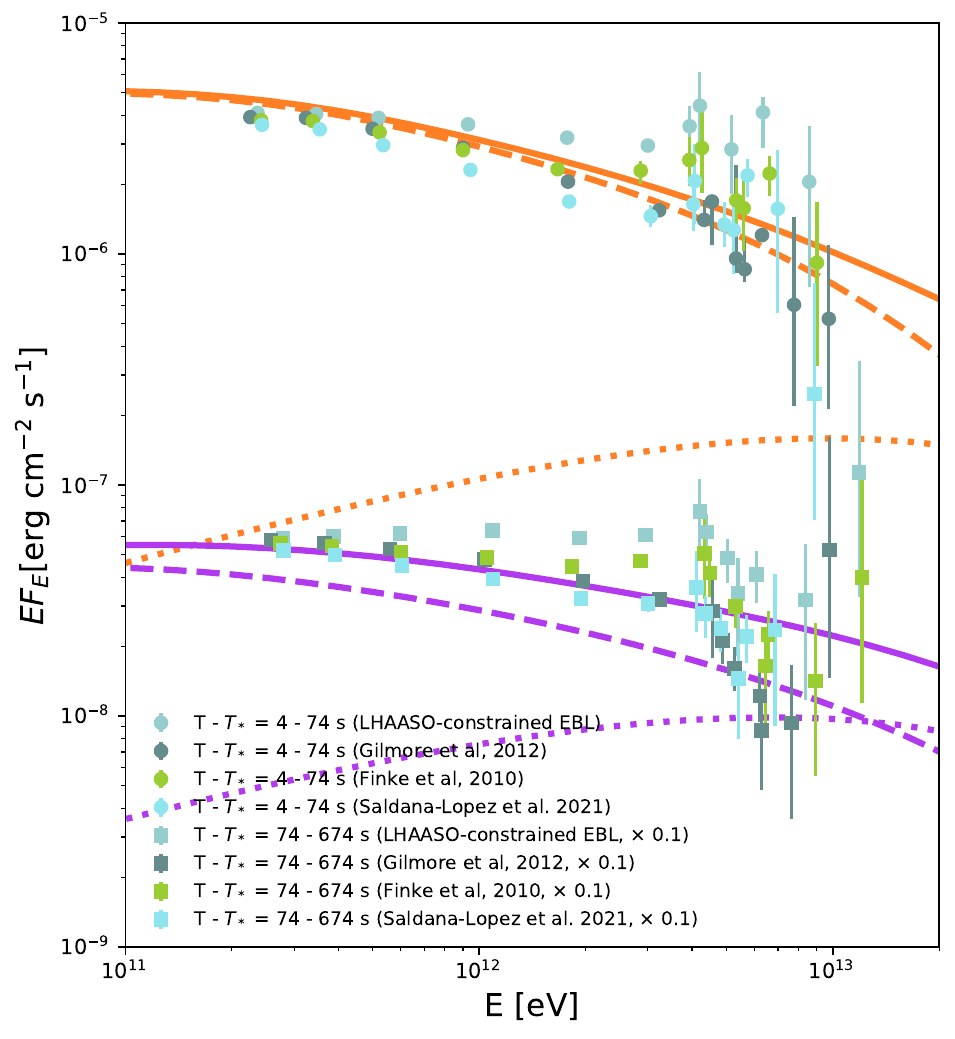}
    \caption{The expected VHE gamma ray energy spectrum from GRB 221009A during the afterglow phase. The dashed lines depict SSC radiation from the forward shock of the narrow core, while dotted lines show proton synchrotron emission from the reverse shock of the wide jet. In the left panel, we compare our theoretical model with the measured energy spectra from LHAASO-WCDA, which go up to 7 TeV. The measurements have been adjusted using an EBL model~\citep{Saldana-Lopez:2020qzx, LHAASO:2023kyg}. The inner bands represent statistical uncertainties, and the outer bands indicate systematic uncertainties. In the right panel, we compare the intrinsic energy spectra after correction for various EBL models, including data from LHAASO-KM2A~\citep{LHAASO:2023lkv}. 
    }
    \label{fig:spectrum}
\end{figure*}

In the lower panel of Fig.~\ref{fig:lightcurve-VHE}, we compare the predicted spectral index evolution with the observed data, where the spectral index is defined as { $EF_E \propto E^{-s+2}$}.
We notice a subtle trend in the spectral index as time progresses~\cite{LHAASO:2023kyg}. Initially, during the rising phase at $T-T_* \sim 10 - 20\rm~s$, it's relatively soft at around $2.4$. As the time elapses, it gradually decreases, reaching a minimum spectral index of approximately $2.2$ at $T-T_* \sim 400\rm~s$. 
Afterward, it starts to become softer again, with a spectral index of around $\sim 2.3$ at $T-T_* \sim 1000\rm~s$.
We observed that the spectral index evolution of the SSC component from the narrow jet does not match the measured values where the spectral index gradually decreases from $2.4$ to $2.3$, as shown by the dashed curve.
Remarkably, the observed change in spectral index finds a straightforward explanation with the introduction of an additional component: the reverse shock proton synchrotron emission.
The energy spectral index of this proton synchrotron emission is $(s_p^r + 1)/2 = 3/2$ for $s_p^r = 2$, which is significantly harder compared to the inverse-Compton spectral index in the TeV energy range~\citep[e.g.,][]{Zhang:2022lff}.
The reduced Chi-square is $\chi^2/{\rm d.o.f.} = 3.3$ by the SSC model curve and it is $\chi^2/{\rm d.o.f.} = 2.8$ with reverse shock proton synchrotron emission as an additional component.
We also performed a Kolmogorov-Smirnov (KS) test to determine which model, SSC or SSC + Psyn, is more consistent with the spectral index evolution. The distribution of normalized residuals for each model is calculated based on the 16 data points shown in the lower panel of Fig.~\ref{fig:lightcurve-VHE}. 
We create a histogram with a range from -3.5 to 3.5 and a bin size of 0.5.
We find the p-value of the KS test with only the SSC component ($1.5\times 10^{-3}$) is smaller than the model including the reverse shock proton synchrotron emission as a second component ($1.1 \times 10^{-2}$), which means the latter gives a better fit to the data. Note the results are affected by the limited number of data points. However, because of the computation cost, this work does not aim to find the best-fit parameters that match the observed light curves and spectra.
We notice that the spectral index's shift towards harder values aligns well with the peak time of the proton synchrotron emission.
Furthermore, this change in spectral behavior correlates with the emergence of a flare at $T-T_* = [320, 550]\rm~s$~\citep{LHAASO:2023kyg}.

In Fig.~\ref{fig:spectrum}, we show the calculated energy spectrum in the early phase ($T - T_* < 2000\rm~s$) of the VHE band, taking into account $\gamma\gamma$ attenuation by prompt photons as in Eq.~\ref{eq:atten_prompt}.
What we see is that the primary source of the observed flux is the SSC emission from the narrow core.
Furthermore, the later spectral hardening observed can be attributed to the reverse shock proton synchrotron emission from the wide jet.

Next, we'll analytically estimate the proton synchrotron emission from the reverse shock at the time of shock crossing, referred to as $T_\times$.
To do this, we assume that the duration of the GRB ejecta, measured in the stellar frame, is $\delta T_{\rm ej} = 100\rm~s$.
The width of the ejecta, measured in the stellar frame, can be approximated as: 
\begin{equation}
\Delta_0 \approx c \delta T_{\rm ej} \simeq 3 \times 10^{12} \delta T_{\rm ej, 2}\rm~cm.
\end{equation}
{
We can estimate the jet spreading radius as
\begin{equation}
R_s \approx \Delta_0 \Gamma_0^2 \simeq 3 \times 10^{17} \delta T_{\rm ej, 2}\rm~cm.
\end{equation}
The deceleration radius is
\begin{equation}
    R_{\rm dec} \simeq 8.4\times 10^{16}\mathcal{E}_{54.5}^{1/3} \Gamma_{0, 2.5}^{-2/3} n_{\rm ex, 2}^{-1/3} \rm~cm.
\end{equation}
Under the condition $R_s > R_{\rm dec}$, the ejecta is in the thick shell case.
}
The radius at which the reverse shock crosses is approximately, 
\begin{equation}
R_\times \simeq 6.4 \times 10^{16} \mathcal{E}_{54.5}^{1/4} \delta T_{\rm ej, 2}^{1/4} n_{\rm ex, 2}^{-1/4} \rm~cm.
\end{equation}
At the time $T_\times = \delta T_{\rm ej}$, the Lorentz factor of the shocked ejecta is
\begin{equation}
\Gamma_\times \simeq 86 \mathcal{E}_{54.5}^{1/8} \delta T_{\rm ej, 2}^{-3/8} n_{\rm ex, 2}^{-1/8}.
\end{equation}
The relative Lorentz factor of the shocked ejecta, when measured in the unshocked ejecta frame, is
\begin{equation}\label{eq:gamma34}
    \Gamma_{34} \simeq 1.9.
\end{equation}
The number density of the unshocked ejecta in the comoving frame can be calculated as
\begin{align}
    n_{\rm ej} &= \frac{\mathcal{E}_k}{4\pi m_p c^2 \Gamma_0 (\Gamma_0 \Delta_0) R_\times^2} \nonumber \\ &= 2.5 \times 10^5 \mathcal{E}_{54.5}^{1/2} \Gamma_{0, 2.5}^{-2} \delta T_{\rm ej, 2}^{-3/2} n_{\rm ex, 2}^{1/2}\rm~cm^{-3}.
\end{align}
The magnetic field strength can be determined as follows, 
\begin{align}
   B_\times &= (32\pi \epsilon_B^r \Gamma_{34} n_{\rm ej}  (\Gamma_{34} - 1) m_p c^2)^{1/2} \nonumber \\ &\simeq 135 \epsilon_{B, -0.5}^{r  1/2} \mathcal{E}_{54.5}^{1/4} \Gamma_{0, 2.5}^{-1} \delta T_{\rm ej, 2}^{-3/4} n_{\rm ex, 2}^{1/4} (g(\Gamma_{34}) / 2.4)^{1/2} \rm~G,
\end{align}
where $g(\Gamma_{34}) \equiv (\Gamma_{34} - 1) (\Gamma_{34} + 3/4)$.
{ In the comoving frame, the acceleration timescale is estimated to be $t_{\rm acc} = \eta \varepsilon_{p, \rm max} / (e B)$, and the dynamical timescale is $t_{\rm dyn} = R_\times / (c \beta \Gamma_\times)$.}
The maximum proton energy { measured in the stellar frame } under the confinement condition $t_{\rm acc} < t_{\rm dyn}$ can be estimated as, 
\begin{align}
E_{p, \rm max}^{\rm dyn} &\approx \Gamma_\times \varepsilon_{p, \rm max} \approx \eta^{-1} \Gamma_\times e B_\times (R_\times / \Gamma_\times) \nonumber \\ &\simeq 2.7 \times 10^{21} \eta^{-1} \epsilon_{B, -0.5}^{r  \frac{1}{2}} \mathcal{E}_{54.5}^{\frac{1}{2}} \Gamma_{0, 2.5}^{-1} \delta T_{\rm ej, 2}^{\frac{-1}{2}} \left(\frac{g(\Gamma_{34})}{2.4}\right)^{\frac{1}{2}} ~{\rm~eV},
\end{align}
where $\eta$ is a coefficient which is a few in the Bohm limit ~\citep[e.g.,][]{Sironi:2015oza}.
The maximum energy is limited by the synchrotron cooling process, and we find, 
\begin{align}
E_{p, \rm max}^{\rm syn} &= \sqrt{\frac{6\pi e}{\sigma_T B_\times \eta Z^3}} \frac{m_p^2 c^2}{m_e} \Gamma_\times \nonumber \\ &\simeq 1.4 \times 10^{21} \eta^{-\frac{1}{2}} Z^{-\frac{3}{2}} \epsilon_{B, -0.5}^{r  -\frac{1}{4}} \nonumber \\ &\times \Gamma_{0, 2.5}^{-\frac{1}{2}}  n_{\rm ex, 2}^{-\frac{1}{4}} \left(\frac{g(\Gamma_{34})}{2.4}\right)^{-\frac{1}{4}} \rm~eV.
\end{align}
The maximum energy of protons is then given by
\begin{equation}
    E_{p, \rm max} = {\rm min} [E_{\rm max}^{\rm dyn}, E_{\rm max}^{\rm syn}].
\end{equation}

The flux of the proton synchrotron emission, which peaks at the maximum energy $E_{\rm psyn} = \Gamma_\times \varepsilon_{\gamma} = {\rm min} [\Gamma_\times \varepsilon_{{\rm psyn}, M}, \Gamma_\times \varepsilon_{{\rm psyn}, c}]$, can be expressed as 
\begin{equation}\label{eq:flux}
    F_E = \frac{1+z}{4\pi d_L^2} N_3 (E_{p, \rm max})  P_E,
\end{equation}
Here, $N_3 (E_{p, \rm max})$ represents the number of protons, and $P_{E}$ is the laboratory frame specific synchrotron power per proton,
\begin{align}\label{eq:power}
    P_E &\approx  \frac{\phi_0 \sqrt{3} e^3}{m_p c^2} B_\times \Gamma_\times \nonumber \\ &\simeq 1.5 \times 10^{-21}\rm~erg~s^{-1}~Hz^{-1},
\end{align}
where $\phi_0$ is a factor of order unity.
The typical energy of the proton synchrotron emission in the observer frame is given by
\begin{align}
    E^{\rm psyn}_{M} &\approx \frac{3}{4\pi} \frac{h e B_\times}{m_p c} \gamma_{p, \rm max}^2 \Gamma_\times  / (1 + z)\nonumber \\ &\simeq 3.5 \times 10^{13} / (1 + z)\rm~eV,
\end{align}
where $\gamma_{p, \rm max} = (E_{p, \rm max} / \Gamma_\times) / m_p c^2$ is the maximum Lorentz factor of protons. 
The cooling energy is determined by the balance between synchrotron cooling timescale and dynamical timescale.  The proton cooling energy is estimated to be $E_{p, c} \simeq 5.6\times 10^{20}\rm~eV$ and the corresponding characteristic photon energy is 
\begin{align}
    E^{\rm psyn}_c &\approx \frac{3}{4\pi} \frac{h e B_\times}{m_p c} \gamma_{p, c}^2 \Gamma_\times  / (1 + z)\nonumber \\ &\simeq 5.8 \times 10^{12} / (1 + z)\rm~eV,
\end{align}
where $\gamma_{p, c} = (E_{p, c} / \Gamma_\times) / m_p c^2$ is the cooling Lorentz factor of protons.
It is important to note that the peak of the proton synchrotron emission in the $E F_E$ spectrum occurs at ${\rm min}[E^{\rm psyn}_c,E^{\rm psyn}_{M}]$.

The comoving frame energy density of nonthermal protons can be calculated as 
\begin{equation}
    \int_{\varepsilon_{p, \rm min}}^{\varepsilon_{p, \rm max}} d\varepsilon_p \varepsilon_p \frac{dn}{d\varepsilon_p} = \epsilon_p^r e_3 = 4 \epsilon_p^r \Gamma_{34}  (\Gamma_{34} - 1)  n_{\rm ej} m_p c^2,
\end{equation}
where $\varepsilon_{p, \rm min} \sim \Gamma_{34} m_p c^2$.
The total number density of nonthermal protons is given by
\begin{equation}
    n_{p, 3} = \int_{\varepsilon_{p, \rm min}}^{\varepsilon_{p, \rm max}} d\varepsilon_p \frac{dn}{d\varepsilon_p}.
\end{equation}
The number of protons at $E_{p, \rm max}$ can be estimated as 
\begin{align}
    N_3 (E_{p, \rm max}) &\approx 4\pi R_\times^2 \Delta_\times \varepsilon_{p} \frac{dn}{d\varepsilon_{p}}|_{\varepsilon_p = \varepsilon_{p, \rm max}} \nonumber \\ &\approx 4\pi R_\times^2 \Delta_\times \frac{\epsilon_p^r 4\Gamma_{34}  (\Gamma_{34} - 1)  n_{\rm ej} m_p c^2}{\rm{ln}(\varepsilon_{p, \rm max} / \varepsilon_{p, \rm min}) \varepsilon_{p, \rm max}} \nonumber \\ &\simeq 1.5 \times 10^{43},
\end{align}
assuming $s_p^r = 2$ and $\epsilon_p^r = 0.6$.
Finally, we can derive the differential flux estimated at $E_{\rm m}$ as
\begin{equation}
    E F_E |_{E = E_{\rm m}} \simeq 3.2 \times 10^{-6} \rm~erg~cm^{-2}~s^{-1}.
\end{equation}
We can see the estimated flux is consistent with Fig.~\ref{fig:lightcurve-VHE} and Fig.~\ref{fig:spectrum}.
Note that the above analytical estimates neglect adiabatic cooling of accelerated protons and internal $\gamma\gamma$ attenuation, which are considered in the numerical calculations. 
Considering that adiabatic cooling reduces the proton synchrotron flux by approximately a factor of 2, and gamma-rays with energies $\gtrsim 10~\rm TeV$ are further suppressed by an additional factor of a few due to internal $\gamma\gamma$ attenuation.

\subsection{Late-time afterglow and reverse shock emission}\label{sec:late-time}
\begin{figure*}
    \centering
    \includegraphics[width=\textwidth]{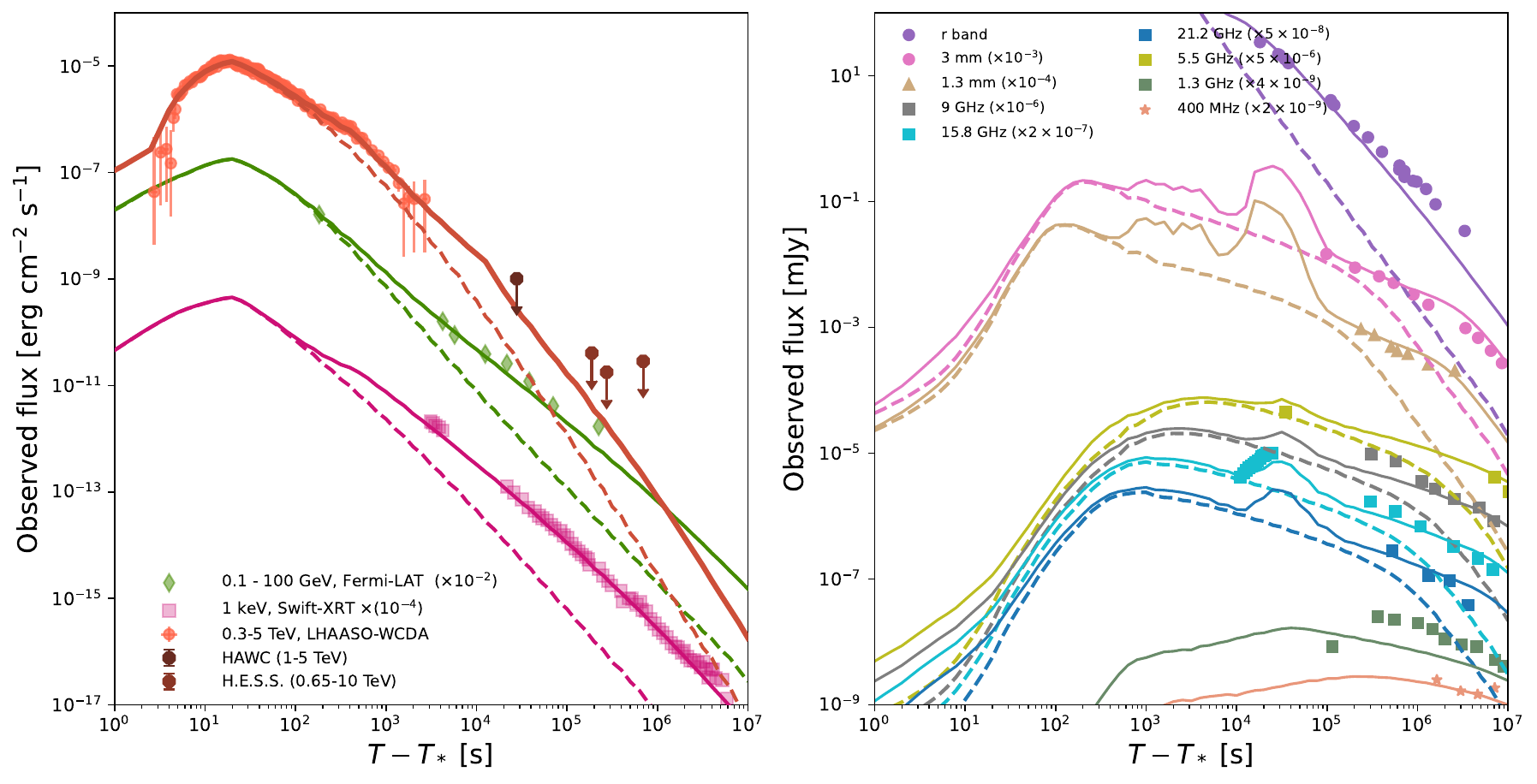}
    \caption{Predicted afterglow light curve for GRB 221009A across multiple wavelengths. The solid lines represent the combined emission from both the narrow core and wide jet, while the dashed lines depict the contributions from the narrow core alone. 
    Light curves for radio, millimeter, optical, and X-ray are taken from~\citep{Laskar:2023yap, OConnor:2023ieu, Bright:2023izk}. The optical data is corrected for extinction. 
    The \textit{Fermi}-LAT light curve is obtained from~\citep{Bissaldi:2023gi}, and the TeV light curve is obtained from~\citep{LHAASO:2023kyg}. 
    Upper limits in the VHE band from H.E.S.S.~\citep{HESS:2023qhy} and HAWC~\citep{2022GCN.32683....1A} are included.}
    \label{fig:lightcurve}
\end{figure*}

In Fig.~\ref{fig:lightcurve}, we compare the predicted multi-wavelength light curve with the observed flux across a wide range of energy bands, from radio to VHE. 

Early on, roughly within the first $10^3\rm~s$, the energy flux is mainly attributed to the external forward shock emission from the narrow jet.
However, at later times, beyond approximately $10^3 - 10^4\rm~s$, the dominant contribution to the energy flux shifts to the external forward shock emission from the wide jet. This transition is highlighted in the left panel of Fig.~\ref{fig:lightcurve}.

In the right panel, we further analyze the energy flux evolution in the lower energy range compared to the observed flux, spanning from radio to optical bands.
Around $T - T_* = 10^4\rm~s$, the energy flux in the optical bands is primarily governed by the external forward shock synchrotron emission from the wide jet.
However, in the radio band, the external forward shock synchrotron emission from the narrow core dominates the energy flux until around $T - T_* = 10^4\rm~s$, when the wide jet's emission takes over. 

\begin{figure}
    \centering
    \includegraphics[width=0.5\textwidth]{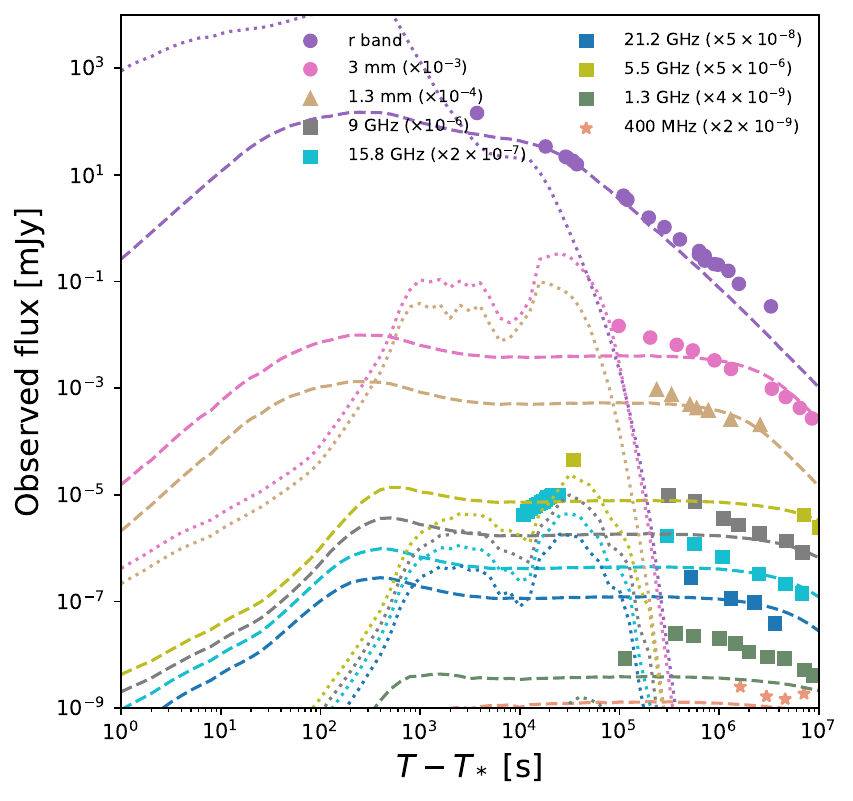}
    \caption{Similar to Fig.~\ref{fig:lightcurve}, we compare the reverse-forward shock emission from the wide jet to the observed data in the low-energy radio to optical bands.  
    The dashed lines represent the external forward shock emission, while the dotted lines represent the external reverse shock emission.
    }
    \label{fig:lightcurve-RS}
\end{figure}

In Fig.~\ref{fig:lightcurve-RS}, we compare the reverse shock electron synchrotron emission from the wide jet to the observed data in the low-energy radio to optical bands. 
Although it's not as prominent as the forward shock emission, 
unlike the reverse shock emission, the forward shock light curve is expected to remain relatively constant in the case of a wind medium, which aligns with the findings presented in ~\cite{Gill:2023ijr}. 
They considered emission from a shallow angular structured jet similar to our wide jet component in the thick shell scenario~\citep{Gill:2023ijr}. 
Similar to ~\cite{Gill:2023ijr, Zheng:2023crd, Ren:2023gdu}, we found that the reverse shock emission from the wide jet is considerably higher than the forward shock emission at $T \sim 10^4\rm~s$ which could account for the observed bump at $\sim 10 - 20\rm~GHz$~\citep{Bright:2023izk, Gill:2023ijr, Zheng:2023crd, Ren:2023gdu}.
Note the two bumps on the reverse shock emission are related to the duration of the ejecta, which is $\delta T_{\rm ej} = 400 \rm~s$ for $\theta < 1.1^\circ$ and $\delta T_{\rm ej} = 3000 \rm~s$ for $\theta > 1.1^\circ$.
It's important to note that photons from the reverse shock electron synchrotron emission play a significant role as they can impact the escape of VHE photons above 10 TeV, which are generated through proton synchrotron emission. %

\section{Discussion and implications}
\label{sec:discuss}

\subsection{VHE gamma rays from GRBs}
In our analysis, we thoroughly investigate the two primary sources of VHE gamma rays in GRB 221009A: SSC emission and proton synchrotron emission. 
Here's what we found: 
\begin{itemize}
    \item For the SSC process, our calculations reveal that narrow-core SSC emission dominates the early TeV afterglow, while the wide-jet SSC emission takes over in the later stages.

    \item Proton synchrotron emission becomes most significant during the transition phase around $T - T_0 \sim 600\rm~s$, where the reverse shock finishes crossing the ejecta of the wide jet. 
    Although the proton synchrotron component doesn't dominate the entire VHE band light curve, it plays a crucial role during this critical period.
\end{itemize}
A closer examination of the VHE spectral index reveals an intriguing pattern: the VHE spectra start soft, then become hard, and finally steepen in the late stages. 
This evolution trend is challenging to explain within the standard SSC model alone. 
It seems more natural to introduce an additional component responsible for spectral hardening.
The proton synchrotron component offers a plausible explanation because it inherently produces harder spectral indices, 
see Sec.~\ref{sec:caveats} for caveats. 
Note that similar behavior of spectral hardening at late time has also been observed in GRB 190829A, which is difficult to explain within the standard SSC model~\citep{Abdalla:2019dlr, Huang:2023dvb}. The reverse shock proton synchrotron component discussed in this work could be helpful to alleviate the limitation of the standard SSC model.

In this work, we employ a two-component jet model to explain the VHE gamma rays observed by LHAASO. 
As demonstrated in Section~\ref{sec:late-time}, the wide-jet component is necessary to account for the late-time radio to optical data. Specifically, the reverse shock radio emission from the wide-jet component aligns with the observed optically thick rising radio light curve.  

The model parameters we found for the wide-jet component are very similar to those adopted in ~Ref.~\cite{Gill:2023ijr}. For example, the isotropic equivalent kinetic energy $\mathcal{E}_k$, $A*$, and $\epsilon_B$ are around 2 times larger than those in ~Ref.~\cite{Gill:2023ijr}. 
Our value of $\epsilon_e$ is 2 times smaller. The power-law index of the energy angular profile $a = 2$ is larger than the value $a \sim 0.8$ found in ~Ref.~\cite{Gill:2023ijr}.
However, Ref.~\cite{Gill:2023ijr} only conducted fitting to the late-time radio to X-ray band, aligning with our results. 
Based on Fig.\ref{fig:lightcurve-VHE}, we found that the SSC emission from the wide-jet (dotted-dashed line) cannot explain the TeV light curve, contributing significantly only at late time $T - T_0 \sim 1000$ seconds. Thus, the observed TeV light curve needs to be explained with another jet component, e.g., the narrow-core~\cite[e.g.,][]{LHAASO:2023kyg}.
In comparison to the model parameters adopted in ~Ref.~\cite{LHAASO:2023kyg}, the width of the narrow core is slightly narrower, with $\theta_j \sim 0.22$ degrees. This narrower jet is necessary to avoid overshooting the late-time radio data. Our calculations suggest that the change in the decaying slope of the light curve is attributed to the emergence of the wide-jet component.

Both work from Ref.~\cite{Zheng:2023crd} and Ref.~\cite{Ren:2023gdu} performed multiwavelength fitting up to the TeV band for GRB 221009A, considering a two-component structured jet model. However, both the model details and the values of physical parameters differ. The isotropic equivalent energy of the wide-jet component we found is over $\sim 1$ order of magnitude more energetic than Ref.~\cite{Zheng:2023crd}, but consistent with Ref.~\cite{Ren:2023gdu}. Compared to Ref.~\cite{Zheng:2023crd} and Ref.~\cite{Ren:2023gdu}, our model is more consistent with the light curve at $100\rm~GHz$ band. We found that the observed X-ray flux around 3000 seconds is dominated by the wide-jet component, which is underpredicted in Ref.~\cite{Ren:2023gdu}. Finally, we should note that our model involves the reverse shock proton synchrotron emission in addition to the SSC component, providing a better description of the TeV light curve and energy spectrum measured by LHAASO.

\subsection{Remarks on proton synchrotron radiation}\label{sec:caveats}

Proton synchrotron emission has been proposed as one of the mechanisms to produce high-energy emission from GRBs~\cite{Totani:1998xb, Totani:1998zr, Zhang:2001az, Kumar:2014upa}. The maximum photon energy produced by high-energy electrons is $E_{\rm syn, e} \sim 50\Gamma/(1+z)\rm~MeV$, while it is $E_{\rm syn, p} \sim 10^2\Gamma/(1+z)\rm~GeV$ for protons.
Given the typical Lorentz factor of the outflow, $\Gamma \sim 100$, the maximum energy of proton synchrotron emission could reach $E_{\rm syn, p} \sim 10\Gamma_2 /(1+z)\rm~TeV$. 
High-energy electron SSC radiation provides a natural explanation for VHE gamma rays, where low-energy seed photons are scattered to the VHE energy range~\cite{Meszaros:1994sd, Dermer:1999eh, Sari:2000zp, Zhang:2001az}.
However, SSC radiation undergoes Klein-Nishina suppression, which makes it challenging to explain the gamma ray spectrum above 10~TeV and the hard spectra index. Thus, proton synchrotron emission are considered as one of the possible mechanisms to produce VHE gamma rays.

For a given energy, the energy loss rate for protons is smaller by a factor of $(m_e/m_p)^4$ than for electrons, where $m_e$ is electron mass and $m_p$ is proton mass~\cite{zhangPhysicsGammaRayBursts2018}. Due to the low efficiency, proton synchrotron emission may require an unreasonably large luminosity to explain \textit{Fermi}-LAT GRBs above $100\rm~MeV$~\cite{Crumley:2012ra}. 
Explaining TeV light curves is less demanding but a large fraction of energy needs to be carried nonthermal protons with $\epsilon_p \lesssim 1$ considering forward shock proton synchrotron emission~\cite{Isravel:2022glo, Isravel:2023thi}.

The reverse shock proton synchrotron emission undergoes a similar issue mentioned above but with the advantage of stronger magnetic fields. From Eqs.~(\ref{eq:flux}) and (\ref{eq:power}), we see that the proton synchrotron emission flux prefers higher magnetic field strength and a larger number of high-energy protons. We adopt a spectral index of $s_p^r = 2$ for protons accelerated by the reverse shock. Our results indicate that $\epsilon_p^r\sim30$\% of ejecta internal energy is transferred to nonthermal protons and $\epsilon_B^r \sim60$\% is converted to magnetic energy.
Note there is parameter degeneracy, if the magnetic field is stronger, the energy fraction by protons is smaller.
Our model also required the acceleration of protons to be very efficient near the Bohm limit with $\eta \sim 1-10$.
We can see the required parameter range is different from the standard ones~\cite{Kumar:2014upa}.

In this work, we simply adopt the basic requirement, $\epsilon_p^r + \epsilon_B^r + \epsilon_e^r \lesssim 1$, to constrain the proton synchrotron emission in our model. 
We consider weakly magnetized shocks in matter-dominated jets, where particle acceleration and the generation of magnetic fields could be efficient~\cite{Sironi:2015oza}, and rather optimistic values of these parameters are still allowed.
For nonrelativistic collisionless shocks, $\epsilon_p^r \sim 10-20$\% is inferred from hybrid particle-in-cell (PIC) simulations~\citep{Caprioli:2013dca}. 
Ref.~\cite{Crumley:2018kvf} conducted kinetic simulations of mildly relativistic shocks and discovered a linear increase in energy consistent with Bohm scaling, and exceeds the rate at ultrarelativistic shocks~\citep{2013ApJ...771...54S}.
Note the PIC simulations are only available for a limited time and length scales, and the long-term evolution is necessary to unveil details~\citep{Groselj:2024dnv}.
The value of $\epsilon_B^r$ depends on the properties of the upstream. The upstream may have $\sigma_{\rm mag,ej} < 1$ where $\sigma_{\rm mag, ej}$ is the ejecta magnetization, but after the shock compression, the value of $\epsilon_B^r$ could be greater than 0.1 while the ejecta is still matter-dominated~\cite{Zhang:2004ie}.
A larger $\epsilon_B^r$ in the reverse shock region than the forward shock region has been revealed from modeling of some previous GRBs such as GRB 990123 (see Ref.~\cite{Zhang:2003wj}) and other GRBs that showed a dominant RS emission in the optical band~\cite{Yi:2020cab}.

Here, we discuss several situations where the extreme values of the parameters mentioned above could be alleviated.
Recent studies have revealed that as a reverse shock moves through a magnetized jet with a large-scale cylindrical magnetic field, particles undergo curvature drift and escape upon reaching the confinement limit. This process leads to a significantly harder spectral index of $s \sim 1$~\citep[e.g.,][]{Huang:2023dqm}. 
By adopting such a hard spectral index of $s_p^r = 1$, we find that converting $\sim10$\% of ejecta internal energy into nonthermal protons with an equal amount into the magnetic field energy is sufficient to explain the observed flux of VHE gamma rays above 10~TeV, potentially alleviating the efficiency issue.
In addition, we assume microphysical parameters such as $\epsilon_B^r$ are homogeneous in the shocked region, but the inhomogeneity in both number density and magnetic field strength can be relevant and could effectively enhance the synchrotron radiation power~\citep{Groselj:2024dnv}. Due to such model uncertainties, the final flux could vary by a factor of 2. Given this uncertainty,  optimistic values of the parameters are still possible, and future high-resolution simulations incorporating long-term evolution are necessary to further explore the feasibility of efficient acceleration, especially in the context of mildly relativistic, magnetized, electron-ion shocks.

Although proton synchrotron emission requires parameters
different from the usual ones, these parameters are still allowed in
this event as mentioned above, and hence the proton synchrotron model
is undoubtedly an interesting possibility for this event.
In the future, high-resolution simultaneous observations of the energy spectrum from the GeV to the TeV band are essential to constrain the peak energy and curvature of the SSC spectrum, and a hard spectrum above the peak energy of the SSC spectrum supports proton synchrotron emission.

\vspace{1em}
\subsection{Ultrahigh-energy neutrinos from GRBs}
In this section, we present the expected flux of high-energy neutrinos from the reverse shock proton synchrotron model. 
The GRB afterglow is expected to emit PeV-EeV neutrinos via the photomeson production process~\citep{Waxman:1999ai, Dermer:2000yd, Li:2002dw, Murase:2007yt, Razzaque:2013dsa}.
However, there is no evidence of a correlation between neutrinos and GRBs analyzed with candidate muon-neutrino events observed by IceCube ~\citep{IceCube:2022rlk, Lucarelli:2022ush}.  
\cite{IceCube:2022rlk} found the total contributions to the quasi-diffuse neutrino flux is less than $24\%$ for an emission timescale $10^4$ seconds.
The detection of PeV-EeV neutrinos would require future UHE neutrino telescopes such as GRAND 200k~\citep{GRAND:2018iaj}, IceCube-Gen2~\citep{IceCube-Gen2:2021rkf}, RNO-G~\citep{RNO-G:2020rmc}, as well as POEMMA~\citep{Venters:2019xwi}.

\begin{figure}
    \centering
    \includegraphics[width=0.5\textwidth]{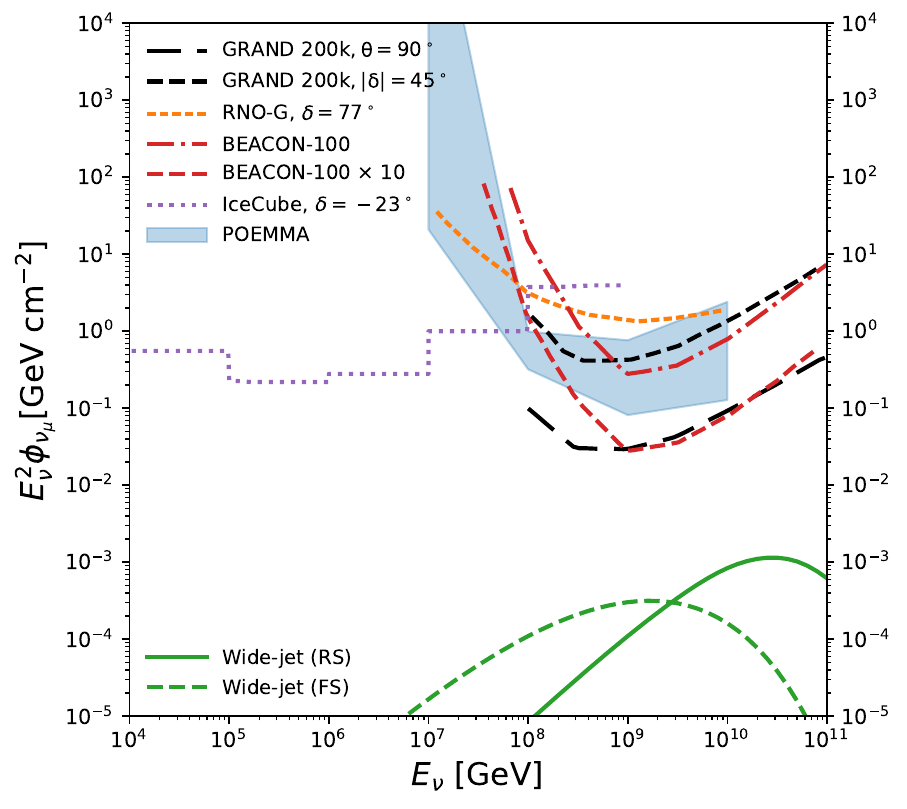}
    \caption{Neutrino fluence predicted in the reverse shock proton synchrotron model for wide-jet. For comparison purposes, we also present the neutrino fluence expected in the forward shock model, assuming a fraction of $\epsilon_p = 0.1$ of the post-shock thermal energy goes into high-energy protons.
    We show the sensitivity of GRAND for a source at declination $\delta=45^\circ$ and  zenith angle $\theta=90^\circ$~\citep{GRAND:2018iaj}, RNO-G at $\delta=77^\circ$~\citep{RNO-G:2020rmc}, 
    IceCube at $\delta = -23^\circ$~\citep{IceCube-Gen2:2021rkf},
BEACON~\citep{Zeolla:2023iqq}, and POEMMA taking the $90\%$
unified confidence level and assuming observations during
astronomical night~\citep{Venters:2019xwi}.
}
    \label{fig:neutrino-fluence}
\end{figure}

The neutrino fluence can be approximated as~\citep[e.g.,][]{Murase:2022vqf, Kimura:2022zyg}
\begin{align}
    E_\nu^2 \phi_{\nu_\mu} &\approx  \frac{1}{8} \frac{1+z}{4\pi d_L^2} {\rm min} [1, f_{p\gamma}] f_{\rm sup} \mathcal{E}_{\rm cr} \frac{1}{\mathcal{R}_{\rm cr}} \nonumber \\ &\simeq { 5.4 \times 10^{-1}} {\rm min} [1, f_{p\gamma}] f_{\rm sup} \rm~GeV~cm^{-2},
\end{align}
where
$f_{p\gamma}$ is the effective optical depth of the photomeson production process with the reverse shock synchrotron emission as the main target photon fields, 
$f_{\rm sup}$ is the suppression factor by pion and muon cooling, $\mathcal{E}_{\rm cr} = \epsilon_p^r 4\pi R_\times^2 \Delta_\times e_3 \Gamma_\times \sim 8 \times 10^{54}\rm~erg$ is the isotropic-equivalent cosmic ray proton energy accelerated by reverse shock, and $\mathcal{R}_{\rm cr} = \rm{ln}(\varepsilon_{p, \rm max} / \varepsilon_{p, \rm min}) \simeq 23$ converts the bolometric cosmic ray energy to the differential cosmic ray energy.
In Fig.~\ref{fig:neutrino-fluence}, we show the predicted single flavor neutrino fluence in the reverse shock proton synchrotron model, which are numerically calculated with Eq.~\ref{eq:flux-EATS} by integrating over the emission time with  $T \sim 10^7$ seconds. 
At the peak energy, the corresponding photomeson production efficiency is estimated to be $f_{p\gamma} \lesssim 0.1$ at the shock crossing time.
For comparison purposes, we also present the neutrino fluence expected in the forward shock model, assuming a fraction of $\epsilon_p = 0.1$ of the post-shock thermal energy goes into high-energy protons and protons could be accelerated to ultrahigh energies.
The accelerated protons follow a power-law distribution with spectral index $s_p = 2$. We can see that the predicted neutrino fluence peaked at the ultra-high-energy range and is lower than the 90\% C.L. upper limits by IceCube.
Even though the neutrino flux from a single GRB is undetectable, the diffuse neutrino flux contributed from all GRBs could provide valuable tests of the model~\citep{Murase:2007yt}.

\subsection{GRBs as UHECR sources}
In this study, we examined proton synchrotron emission from protons accelerated to ultra-high energies. 
We estimated that the total energy of these accelerated nonthermal protons is approximately
{ $\mathcal{E}_{\rm cr} \sim 8 \times 10^{54}\rm~erg$}.
However, detecting escaping UHECR protons from GRBs simultaneously with VHE gamma rays is challenging. 
UHECR undergoes various energy loss processes during the journey to Earth~\citep[e.g.,][]{Waxman:1995vg, Vietri:1995gn}.
The influence of magnetic fields within both host galaxies and extragalactic space can not be neglected. These magnetic fields can introduce significant time delays in the arrival of UHECRs compared to the rectilinear propagation scenario~\citep[e.g.,][]{Miralda-Escude:1996twc, Murase:2008sa}. 
Note that the magnetic field in our galaxy also deflects the arrival direction of UHECRs and causes significant time delays, as shown in \cite{Murase:2008sa}.
A recent study suggested that a UHECR burst from GRB 221009A could be detected within about 10 years if inter-galactic magnetic fields with strength $\lesssim 10^{-13}\rm~G$~\citep{2024arXiv240111566H}. 
However, in our scenario, we consider UHECRs escaping directly from the source, with neutron escape playing an insignificant role. Consequently, we predict a lower flux of UHECRs from GRBs compared to the estimates in \citet{2024arXiv240111566H}. Furthermore, the combined effects of magnetic fields within galaxies and large-scale structures~\citep{Takami:2011nn}, and energy losses during intergalactic propagation significantly complicate the detection of such emissions.
Additionally, the UHECR-induced intergalactic electromagnetic cascades could also generate VHE gamma rays~\citep{AlvesBatista:2022kpg, Das:2022gon, Mirabal:2022ipw}.
However, the time delay and the development of the electromagnetic cascades strongly depend on the structure and strength of intergalactic magnetic fields~\citep{2012ApJ...749...63M}, which require further investigation.

\section{Summary}
\label{sec:summary}
In this study, we conducted a comprehensive investigation of radiation within the standard external reverse-forward shock scenario, considering a two-component structured jet comprising a narrow core dominated by magnetic energy and a wide jet dominated by matter.
We compared our numerical findings with observed energy spectra and multi-wavelength light curves spanning from radio to VHE bands. 
Our key results are summarized as follows:
\begin{itemize}
    \item The energy flux light curve observed in the VHE band is primarily dominated by SSC emission from the narrow core. The SSC emission from the wide jet only becomes dominant at late times, approximately beyond $10^3$ seconds. 
    Reverse shock proton synchrotron emission becomes significant when the reverse shock crosses the shocked ejecta of the wide jet around $T - T_0 \sim 600$ seconds, coinciding with the emergence of a mild flare around $T - T_0 \sim [546, 776]$ seconds after the trigger.

    \item The initial rapid rise of the VHE band light curve is due to $\gamma\gamma$ attenuation caused by prompt target photons
    , although the reference time $T_*$ is uncertain, and a different $T_*$ changes the early light curve significantly.

    \item Compared to the inverse-Compton process, reverse shock proton synchrotron emission exhibits a much harder spectrum in the VHE band. This naturally explains the observed energy spectrum hardening and potentially contributes significantly to the photon flux above $\gtrsim 10\rm~TeV$.

    \item The multi-wavelength afterglow light curve from radio to the GeV band can be explained within the two-component structured jet model. The emission from the wide jet starts to dominate the observed multi-wavelength afterglow after $T - T_0 \sim 10^3 - 10^4$ seconds, while radio emission from the narrow core may continue to dominate the flux at even later times, until around $T - T_0 \sim 10^4$ seconds. The reverse shock emission from the wide jet could explain the steep rise of the $\sim 10 \rm~GHz$ radio data at $T - T_0 \sim 10^4$ seconds.

    \item Our findings may support the idea that GRBs are efficient accelerators of UHECRs, especially at the reverse shock. Further multimessenger observations involving more events are essential to better understand the energy budget of UHECRs originating from GRBs.
\end{itemize}

\section*{acknowledgments}
The work of K.M. is supported by the NSF grants Nos.~AST-2108466, AST-2108467, and AST-2308021. It is also supported by KAKENHI Nos.~20H01901 and 20H05852 (B.T.Z. and K.M.), and No.~22H00130, 20H01901, 20H00158, 23H05430, 23H04900 (K.I.). 
We acknowledge support by Institut Pascal at Universit\'{e} Paris-Saclay during the Paris-Saclay Astroparticle Symposium 2023, with the support of the P2IO Laboratory of Excellence (program “Investissements d’avenir” ANR-11-IDEX-0003-01 Paris-Saclay and ANR-10-LABX-0038), the P2I axis of the Graduate School of Physics of Université Paris-Saclay, as well as IJCLab, CEA, IAS, OSUPS, and the IN2P3 master projet UCMN (B.T.Z.). 

\appendix

\section{Numerical modeling of nonthermal radiation from the external reverse-forward shock}\label{eq:appendix_numerical}
\subsection{Dynamics}
\label{sec:dynamic}
\begin{figure}
    \centering
    \includegraphics[width=0.5\textwidth]{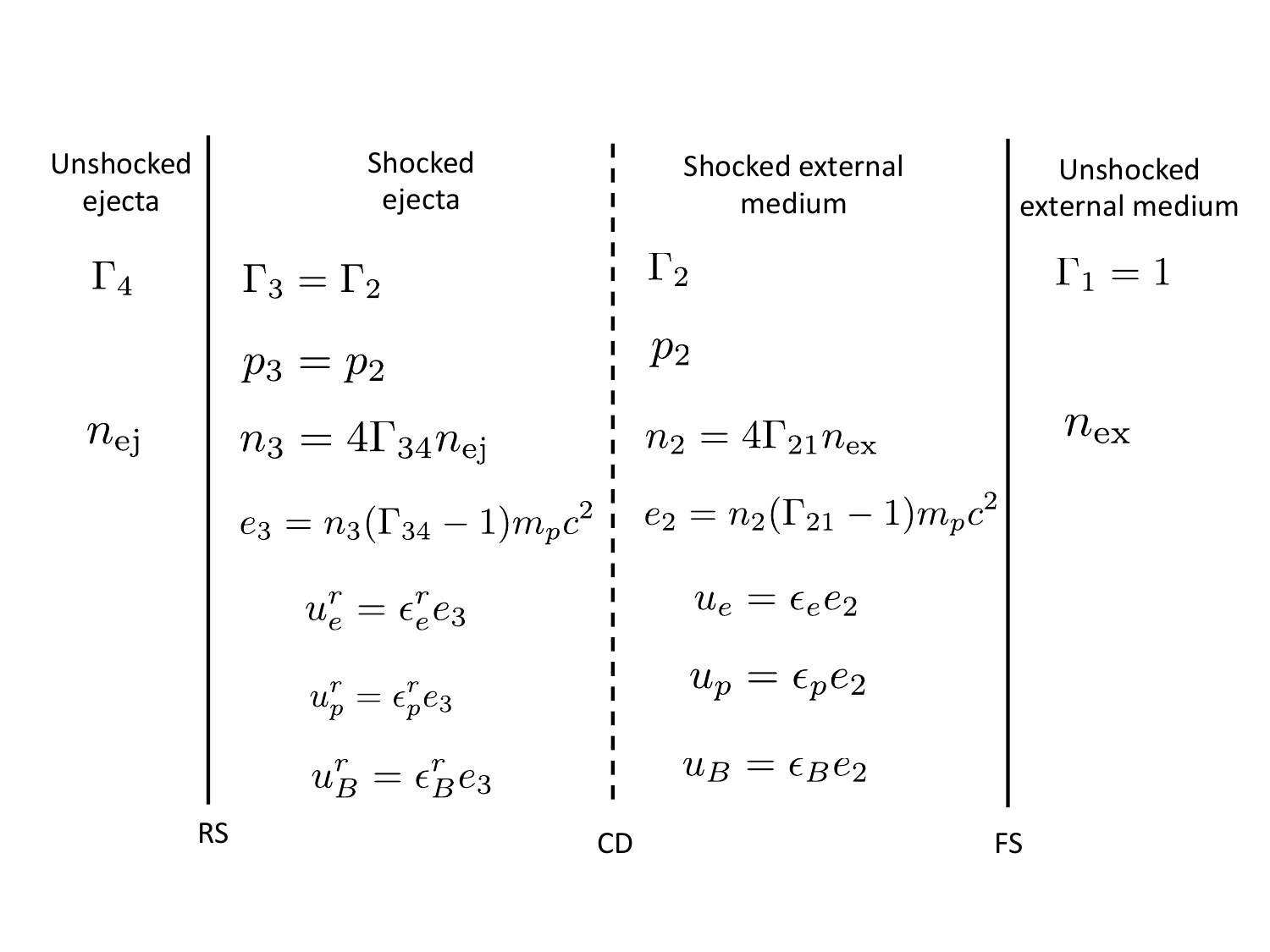}
    \caption{{ Notations and shock jump conditions} for the reverse-forward shock system.}
    \label{fig:RSFS}
\end{figure}

When ultra-relativistic ejecta of the GRB propagate into the external medium, the ejecta are decelerated as their kinetic energy is transferred to the external medium. Two shocks are formed: a forward shock advancing into the ambient external medium, and a reverse shock moving backward in the ejecta rest frame, as illustrated in Fig.~\ref{fig:RSFS}.

Following the method proposed in ~\cite{Nava:2012hq, zhangPhysicsGammaRayBursts2018}, we incorporate the influence of the reverse shock on the dynamical evolution of the outflow.
We solve the following differential equations,
\begin{align}
    \frac{d\Gamma}{dR} &= -\frac{(\Gamma_{\rm eff,2} + 1)(\Gamma - 1) c^2 \frac{dm}{dR} + \Gamma_{\rm eff,2} \frac{d\mathcal{E}_{\rm ad,2}^\prime}{dR}}{(M_{\rm ej,3} + m)c^2 + \mathcal{E}_{\rm int,2}^\prime \frac{d\Gamma_{\rm eff,2}}{d\Gamma} + \mathcal{E}_{\rm int,3}^\prime \frac{d\Gamma_{\rm eff,3}}{d\Gamma}} \nonumber \\ &- \frac{(\Gamma - \Gamma_0 - \Gamma_{\rm eff, 3} + \Gamma_{\rm eff, 3} \gamma_{34}) c^2\frac{dM_{\rm ej,3}}{dR} + \Gamma_{\rm eff, 3} \frac{d\mathcal{E}_{\rm ad,3}^\prime}{dR}}{(M_{\rm ej,3} + m)c^2 + \mathcal{E}_{\rm int,2}^\prime \frac{d\Gamma_{\rm eff,2}}{d\Gamma} + \mathcal{E}_{\rm int,3}^\prime \frac{d\Gamma_{\rm eff,3}}{d\Gamma}},
\end{align}
where 
\begin{equation}
\Gamma_{\rm eff, 2} \equiv (\hat{\gamma}\Gamma^2 - \hat{\gamma} + 1)/\Gamma,
\end{equation}
in needed to properly describe the Lorentz transformation of the internal energy, and the corresponding adiabatic index is $\hat{\gamma} = (4\Gamma + 1) / 3\Gamma$.
When calculating $\Gamma_{\rm eff, 3}$, we use $\hat{\gamma} = (4\Gamma_{\rm 34} + 1) / 3\Gamma_{\rm 34}$ where $\Gamma_{34} = \Gamma \Gamma_0 (1 - \beta \beta_0)$ represents the relative Lorentz factor between the shocked and unshocked ejecta. Note parameters with $\prime$ represent comoving frame value.

The mass of the unshocked ejecta that completes crossing the reverse shock is given by:
\begin{equation}
    \frac{dM_{\rm ej,3}}{dR} = \frac{M_{\rm ej}}{c T_{\rm ej}} \frac{\beta_0 - \beta}{\beta}.
\end{equation}
The evolution of the internal energy of the shocked external matter can be determined by the differential equation,
\begin{equation}
    \frac{d\mathcal{E}_{\rm int, 2}^\prime}{dR} = (1 - \epsilon_\gamma) (\Gamma - 1) 4\pi R^2 n_{\rm ex} m_p c^2 + \frac{d\mathcal{E}_{\rm ad, 2}^\prime}{dR},
\end{equation}
where $\epsilon_\gamma$ represents the radiation efficiency.
The evolution of the adiabatic energy of the shocked external medium is described by the equation: 
\begin{equation}
    \frac{d\mathcal{E}_{\rm ad, 2}^\prime}{dR} = -(\hat{\gamma} - 1) \left(\frac{3}{R} - \frac{1}{\Gamma} \frac{d\Gamma}{dR}\right) \mathcal{E}_{\rm int, 2}^\prime.
\end{equation}
The evolution of the internal energy of the shocked ejecta can be determined by the following differential equation
\begin{equation}
    \frac{d\mathcal{E}_{\rm int,3}^\prime}{dR} = (\Gamma_{34} - 1) \frac{dM_{\rm ej,3}}{dR} c^2 + \frac{d\mathcal{E}_{\rm ad, 3}^\prime}{dR}.
\end{equation}
The evolution of the adiabatic energy of the shocked ejecta is given by
\begin{equation}
    \frac{d\mathcal{E}_{\rm ad, 3}^\prime}{dR} = -(\hat{\gamma} - 1) \left(\frac{3}{R} - \frac{1}{\Gamma} \frac{d\Gamma}{dR}\right) \mathcal{E}_{\rm int, 3}^\prime.
\end{equation}
Once the reverse shock finishes crossing the ejecta, the evolution of the shocked ejecta follows the power law scaling $\Gamma \propto R^{-g}$, with $g = 2$ in the thin shell case and $g = 7/2$ in the thick shell case~\citep{2000ApJ...545..807K}.
The evolution of the blastwave radius is given by
\begin{equation}
    \frac{dR}{dt} = \beta c,
\end{equation}
where $\beta = \sqrt{1 - 1 / \Gamma^2}$.
The evolution of the jet angular width is
\begin{equation}
    \frac{d\theta_j}{dR} = \frac{c_s}{\Gamma R c},
\end{equation}
where $c_s$ is the speed of sound.
Note in this work, the { lateral expansion of the jet} is not taken into account~\citep{2009ApJ...698.1261Z}. 

\subsection{Radiative processes}
\subsubsection{Particle acceleration and cooling}
Based on the diffusive shock acceleration mechanism, the accelerated particles, including both electrons and ions, are assumed to follow a power law distribution with an exponential cutoff with the maximum momentum,
\begin{equation}
    \dot{n}_p^{\rm inj} = \mathcal{C} p^{-s} {\rm exp} (-p/p_{\rm max}) \ \ \ (p > p_{\rm inj}),
\end{equation}
where $p_{\rm inj}$ is the comoving frame injection momentum, $p_{\rm max}$ is the comoving frame maximum momentum, $s$ is the spectral index.
The downstream internal energy density in the shocked external medium is
\begin{equation}
    e_2 = (\Gamma_{21} - 1) n_2 m_p c^2 = (\Gamma_{21} - 1) 4\Gamma_{21} n_1 m_p c^2,
\end{equation}
where { $\Gamma_{21} = \Gamma$} is the relative Lorentz factor.
The normalization constant $\mathcal{C}$ is constrained with following equation,
\begin{equation}
    t_{\rm dyn} \int dp \dot{n}_p^{\rm inj} = n_2.
\end{equation}
We also assume particles are injected at a constant rate within the dynamical time.
The comoving frame dynamical timescale is
\begin{equation}
t_{\rm dyn} \approx R / \Gamma \beta c.
\end{equation}
The acceleration timescale is parameterized as 
\begin{equation}
    t_{\rm acc} = \eta t_L = \eta \frac{p}{Z e B c},
\end{equation}
and the maximum acceleration momentum is given by
\begin{equation}
    p_{\rm max, dyn} = \frac{1}{\eta} \frac{R}{\Gamma \beta} Z e B. 
\end{equation}
The synchrotron cooling timescale is
\begin{equation}\label{eq:syn-cooling}
    t_{\rm syn}^{-1} = \frac{\sigma_T Z^4 m_e^2 B^2}{6 \pi m^4 c^3} p. 
\end{equation}
The maximum acceleration is limited by synchrotron cooling
\begin{equation}
    p_{\rm max, syn} = \sqrt{\frac{1}{\eta} \frac{6\pi e}{\sigma_T Z^3 B}} \frac{m^2 c^2}{m_e},
\end{equation}
and we have
\begin{equation}
    p_{\rm max} = {\rm min} [p_{\rm max, dyn}, p_{\rm max, syn}].
\end{equation}

For electrons, the minimum injection momentum of the nonthermal electrons is 
\begin{equation} \label{eq:inj}
    p_{e, \rm inj} = [(\varepsilon_{e, \rm kin} + m_e c^2)^2 - m_e^2 c^4)]^{1/2},
\end{equation}
where
\begin{equation} \label{eq:kin}
    \varepsilon_{e, \rm kin} = \frac{\epsilon_e}{f_e} g(s_e) (\Gamma - 1) m_p c^2,
\end{equation}
and 
\begin{equation}
    g(s_e) \approx \begin{cases}
        \frac{s_e - 2}{s_e - 1}, s_e > 2 \\
        {\rm ln}^{-1} \left(\frac{p_{e, \rm max}}{p_{e, \rm inj}}\right), s_e = 2 \\
    \end{cases}.
\end{equation}
Note that Eq.~\ref{eq:inj}-\ref{eq:kin} are also valid in the nonrelativistic case, which is essential for late-time afterglow modeling.
For protons, the minimum injection energy is assumed to be thermal energy, then the minimum injection momentum is estimated to be
\begin{equation}
    p_{p, \rm inj} \approx [((\Gamma - 1) m_p c^2 + m_p c^2)^2 - m_p^2 c^4]^{1/2} = (\Gamma^2 - 1)^{1/2} m_p c^2.
\end{equation}
The particle distribution function per unit volume, $n_{p} = \partial n / \partial p$, is determined by solving the following kinetic equation~\citep[e.g.,][]{Blumenthal:1970gc, Zhang:2020qbt, Derishev:2021ivd}
\begin{equation}\label{eq:kinetic-equation}
    \frac{n_{p}}{\partial t} = -\frac{\partial}{\partial p} (\dot{p} n_{p}) - \dot{n}_p^{\rm inj} - \frac{n_{p}}{t_{\rm exp}} - \frac{n_{p}}{t_{\rm esc}},
\end{equation}
where 
$\dot{n}_p^{\rm inj}$ is the particle injection rate, $t_{\rm exp}$ is the expansion time of the shocked region,
$t_{\rm esc}$ is the possible escape time,
\begin{equation}
    \dot{p} \equiv \frac{dp}{dt} = p t_{\rm cool}^{-1} = p [t^{-1}_{\rm syn}(p) + t^{-1}_{\rm IC}(p) + t^{-1}_{\rm ad}],
\end{equation}
is the particle cooling rate, including synchrotron energy losses, inverse-Compton cooling, and adiabatic cooling.
We determine the particle distribution with the iteration method~\citep{murase_implications_2011, Zhang:2020qbt},
\begin{equation}\label{eq:iteration-solution}
{ n_{p}} = \frac{1}{{t_{\rm cool}^{-1}} + t
_{\rm dyn}^{-1}}\frac{1}{p}\int_{p_{\rm eff}}^{ \infty} dp^\prime \dot{n}_p^{\rm inj} (p^\prime).
\end{equation}
In Ref.~\cite{Derishev:2021ivd}, the term $1 / t_{\rm eff} = 1 / t_{\rm esc} + 1 / t_{\rm exp}$ is treated as the effective particle lifetime. In this work, we consider the maximum particle lifetime as $t_{\rm dyn}$.  In the slow cooling case, Eq.~\ref{eq:iteration-solution} gives a reasonably good approximation of the particle energy spectrum derived through numerically solving the time-dependent kinetic equation. In the fast-cooling regime, Eq.~\ref{eq:iteration-solution} gives the steady-state solution~\citep[See Eq. C.3 in][]{Dermer:2009zz}.
In addition, we adopt
\begin{equation}
p_{\rm eff} = \frac{p_{\rm { inj}} p_{\rm cool}}{p_{\rm { inj}} + p_{\rm cool}},
\end{equation}
as the effective particle momentum determined by particle injection momentum $p_{\rm inj}$ and cooling momentum $p_{\rm cool}$, which gives a better approximation to the particle distribution when $p_{\rm inj} \sim p_{\rm cool}$.
We can observe that $p_{\rm eff} \approx p_{\rm inj}$ when  $p_{\rm inj} \ll p_{\rm cool}$, and $p_{\rm eff} \approx p_{\rm cool}$ when $p_{\rm inj} \gg p_{\rm cool}$.
For protons, gives that $p_{\rm inj} \ll p_{\rm cool}$, we have $p_{\rm eff} \approx p_{\rm inj}$.
The dynamical timescale, measured in the comoving frame, is defined as
\begin{equation}
    t_{\rm dyn} = \int \frac{dR}{\Gamma \beta c}. 
\end{equation}
Due to the expansion of the emitting region, particles undergo adiabatic energy losses, and the energy lost rate is estimated to be~\cite[e.g.,][]{Dermer:2009zz}
\begin{equation}
    \left|\frac{dp_e}{dR}\right|_{\rm ad} = \frac{p_e}{R} \left(1 - \frac{1}{3} \frac{d {\rm ln}\Gamma}{d {\rm ln}R} \right),
\end{equation}
and the corresponding adiabatic cooling timescale can be estimated as
\begin{equation}
     t_{\rm ad}  = \left(\frac{1}{p_e} \left|\frac{dp_e}{dR}\right|_{\rm ad} \frac{dR}{dt}\right)^{-1} = \frac{R}{\Gamma \beta c} \left(1 - \frac{1}{3} \frac{R}{\Gamma} \frac{d\Gamma}{dR}\right)^{-1}.
\end{equation}
The expansion time of the shocked region is $t_{\rm exp} = V/\dot{V} = R/3\dot{R} = t_{\rm ad} / 3$~\citep{1975ApJ...196..689G, Derishev:2021ivd}.
The synchrotron cooling timescale is given by Eq.~\ref{eq:syn-cooling}.
The inverse-Compton cooling timescale of nonthermal electrons can be approximated as
\begin{equation}
    t_{\rm IC}^{-1} = \frac{1}{\varepsilon_e} \left|\frac{d\varepsilon_e}{dt}\right|_{\rm IC} \approx \frac{4}{3} \sigma_T c \frac{\varepsilon_e}{m_e^2 c^4} u_{\rm ph}(\varepsilon_\gamma) F_{\rm KN},
\end{equation}
where $F_{\rm KN}$ is a factor to take into account the Klein-Nishina effect.

The synchrotron emissivity can be calculated using Eq. C3 in ~\cite{Zhang:2020qbt}. Note there is a typo in Eq. C3, where an additional term $\varepsilon^\prime$ in the denominator is unnecessary. The numerical calculation is not affected.
The formula used in AMES is
\begin{equation}
j_{\varepsilon_\gamma}^{\rm syn} = \frac{\sqrt{3}}{4\pi} \frac{e^3 B}{m_e c^2 2 \pi \hbar} \int d\gamma_e n_{\gamma_e} G(x),
\end{equation}
where
\begin{equation}
G(x)\approx\frac{1.81 e^{-x}}{(x^{-2/3} + (3.62/\pi)^2)^{1/2}},
\end{equation}
$x = \varepsilon_\gamma / \varepsilon_c$, and  $\varepsilon_c = (3e\hbar B / 2 m_e c) \gamma_e^2 \beta^2$ is the critical energy.

The comoving SSC emissivity is calculated with Eq. C5 in ~\cite{Zhang:2020qbt}.
Note the comoving-frame target synchrotron photon density is estimated as
\begin{equation}
    \frac{dn^{\rm syn}}{d\varepsilon_{\gamma}}= 4\pi j_{\varepsilon_\gamma}^{\rm syn} \times t_{\rm esc},
\end{equation}
where $t_{\rm esc} \approx 2 w / c$ is the photon escape timescale and $w$ is the width of the shocked shell estimated in Eq.~\ref{eq:width}.
The factor of $2$ takes into account the geometrical effect when photons escape from a thin shell~\citep{Fukushima:2017hzm, Derishev:2019cgi}.

\subsubsection{Emission from reverse shock}
The total number of protons (or electrons) in the ejecta shell is given by
\begin{equation}
    N_{0, 4} = \frac{\mathcal{E}_k}{\Gamma_0 m_p c^2},
\end{equation}
where $\Gamma_0$ is the initial Lorentz factor and $\mathcal{E}_k$ is the isotropic-equivalent kinetic energy.
The comoving frame number density of the unshocked ejecta is given by
\begin{equation}
    n_{4} (R) = \frac{N_{0, 4}}{4\pi R^2 \Gamma_0 \Delta_0} = \frac{\mathcal{E}_k}{4\pi m_p c^2 \Gamma_0 (\Gamma_0 \Delta_0) R^2},
\end{equation}
where $\Delta_0$ is the initial width of the ultra-relativistic ejecta shell that represents the geometrical thickness measured in the laboratory frame. 
It is defined as
\begin{equation}
    \Delta_0 = {\rm max} \left[c T_{\rm ej} / (1 + z), \frac{R}{2\Gamma_0^2}\right],
\end{equation}
where $R / 2\Gamma_0^2$ represents the shell width due to the radial spreading of the ejecta.

The comoving frame number density of protons (or electrons) in the shocked ejecta is given by
\begin{equation}
    n_{3} = 4\Gamma_{34} n_{4},
\end{equation}
where 
\begin{equation}
    \Gamma_{34} = \frac{1}{2} \left(\frac{\Gamma_0}{\Gamma} + \frac{\Gamma}{\Gamma_0} \right).
\end{equation}

The total number of protons (or electrons) in the shocked ejecta shell can be estimated as
\begin{equation}
    N_3 = N_{0, 4} \frac{M_{\rm ej, 3}}{M_{\rm ej}}.
\end{equation}
The comoving frame width of the shocked ejecta shell can be estimated as
\begin{equation}
    \Delta_3 = \frac{N_3}{4\pi R^2 n_3}.
\end{equation}
Then, we can estimate the value of $\Delta s$ as 
\begin{equation}
    \Delta s = \frac{\Delta_3}{ \Gamma |\mu - \beta_{\rm sh}|}.
\end{equation}
The downstream internal energy density is given by
\begin{equation}
    e_3 = (\Gamma_{34} - 1) n_3 m_p c^2.
\end{equation}
The downstream magnetic field strength is
\begin{equation}
    B_3 = (8\pi \epsilon_{B, r} e_3)^{1/2}.
\end{equation}
The minimum electron energy is
\begin{equation}
    \varepsilon_{m, e} = g(s_{e, r}) \frac{\epsilon_{e, r}}{f_{e, r}} (\Gamma_{34} - 1) \frac{m_p}{m_e} m_e c^2,
\end{equation}
where $g(s_{e, r}) = (s_{e, r}-2)/(s_{e, r}-1)$ for $s_{e, r} > 2$, and $s_{e, r}$ is the nonthermal electron spectral index in the reverse shock region.
The maximum electron energy is
\begin{equation}
    \varepsilon_{M, e} =  \left(\frac{6\pi e}{\sigma_T B_3 \eta_{r}}\right)^{1/2} m_e c^2,
\end{equation}
where $\eta_{r}$ is a parameter that describes the details of acceleration, with a typical value of around a few in the Bohm limit.
Similar to the forward shock, we determine the nonthermal electron distribution with Eq.~\ref{eq:iteration-solution} before the reverse shock finishes crossing the ejecta.

In the post RS-crossing phase, we adopt a parameterized power law decay of the shocked ejecta
\begin{equation}
    \Gamma_3 \propto r^{-g},
\end{equation}
where $g = 7/2$ in the thick shell case and $g = 2$ in the thin shell case~\citep{2000ApJ...545..807K, zhangPhysicsGammaRayBursts2018}.
Note that the total number of protons (or electrons) remains constant,
\begin{equation}
    N_{3} = N_{0, 4}.
\end{equation}

In the thick shell case, 
the evolution of the comoving frame energy density is given by
\begin{equation}
    e_{3} = e_{3} (t_\times) \left(\frac{R}{R_\times}\right)^{-26/3},
\end{equation}
and the number density is
\begin{equation}
    n_{3} = n_{3} (t_\times) \left(\frac{R}{R_\times}\right)^{-13/2}.
\end{equation}
The evolution of the magnetic field is given by
\begin{equation}
    B_{3} = B_{3} (t_\times) \left(\frac{R}{R_\times}\right)^{-13/3}.
\end{equation}
Additionally, the evolution of the comoving frame width is
\begin{equation}
    \Delta_3 = \Delta_3 (t_\times) \left(\frac{R}{R_\times}\right)^{9/2}.
\end{equation}

In the thin shell case, the evolution of the energy density is
\begin{equation}
    e_{3} = e_{3} (t_\times) \left(\frac{R}{R_\times}\right)^{-8(3+g)/7},
\end{equation}
and the number density is
\begin{equation}
    n_{3} = n_{3} (t_\times) \left(\frac{R}{R_\times}\right)^{-6(3+g)/7},
\end{equation}
The evolution of the magnetic field is given by
\begin{equation}
    B_{3} = B_{3} (t_\times) \left(\frac{R}{R_\times}\right)^{-4(3+g)/7}.
\end{equation}
Additionally, the evolution of the comoving frame width is
\begin{equation}
    \Delta_3 = \Delta_3 (t_\times) \left(\frac{R}{R_\times}\right)^{6(3+g)/7 - 2}.
\end{equation}

\bibliographystyle{elsarticle-harv}
\bibliography{main}

\end{document}